\documentclass[usenatbib]{mnras}
\usepackage[utf8]{inputenc}
\usepackage{amsmath}
\usepackage{amssymb}
\usepackage{graphicx}
\usepackage[mathscr]{euscript}
\usepackage[utf8]{inputenc}
\usepackage[T1]{fontenc}
\usepackage{ae,aecompl}
\usepackage{subfig}
\usepackage{float}

\newcommand{\ksk}{$\rm km\ s^{-1}\ kpc^{-1}$}
\newcommand{\kms}{$\rm km\ s^{-1}$}

\title[Fluctuations in bar parameters]{Fluctuations in galactic bar parameters due to bar-spiral interaction}

\author[Hilmi, Minchev et al.]{
T.~Hilmi$^{1,2}$\thanks{E-mail: th00229@surrey.ac.uk},
I.~Minchev$^{1}$\thanks{E-mail: iminchev@aip.de},
T.~Buck$^{1}$,
M.~Martig$^{3}$,
A.~C.~Quillen$^{4}$,
G.~Monari$^{5}$,
\newauthor
B.~Famaey$^{5}$,
R.~S.~de~Jong$^{1}$,
C.~F.~P.~Laporte$^6$,
J.~Read$^{2}$,
J.~L.~Sanders$^7$,
\newauthor
M.~Steinmetz$^{1}$,
C.~Wegg$^{8}$
\\
$^{1}$Leibniz Institut f\"ur Astrophysik Potsdam (AIP), An der Sternwarte 16, D-14482, Potsdam, Germany\\
$^{2}$Astrophysics Research Group, University of Surrey, Guildford, Surrey GU2 7XH, UK\\
$^3$Astrophysics Research Institute, Liverpool John Moores University, 146 Brownlow Hill, Liverpool L3 5RF, UK\\
$^4$Department of Physics and Astronomy, University of Rochester, Rochester, NY 14627\\
$^5$Universit\'e de Strasbourg, CNRS UMR 7550, Observatoire astronomique de Strasbourg, 11 rue de l'Universit\'e, 67000 Strasbourg, France\\
$^6$Kavli IPMU (WPI), UTIAS, The University of Tokyo, Kashiwa, Chiba 277-8583, Japan\\
$^7$Institute of Astronomy, University of Cambridge, Madingley Road, Cambridge CB3 0HA\\
$^8$Universit\'e C\^ote d'Azur, Observatoire de la C\^ote d'Azur, CNRS, Laboratoire Lagrange, France
}

\date{Accepted 26 June 2020}

\pubyear{2020}

\begin{document}
\label{firstpage}
\pagerange{\pageref{firstpage}--\pageref{lastpage}}
\maketitle 

\begin{abstract}
We study the late-time evolution of the central regions of two Milky Way-like simulations of galaxies formed in a cosmological context, one hosting a fast bar and the other a slow one. We find that bar length, $R_{\rm b}$, measurements fluctuate on a dynamical timescale by up to 100\%, depending on the spiral structure strength and measurement threshold. The bar amplitude oscillates by about 15\%, correlating with $R_{\rm b}$. The Tremaine-Weinberg-method estimates of the bars' instantaneous pattern speeds show variations around the mean of up to $\sim20\%$, typically anti-correlating with the bar length and strength. Through power spectrum analyses, we establish that these bar pulsations, with a period in the range $\sim60-200$~Myr, result from its interaction with multiple spiral modes, which are coupled with the bar. Because of the presence of odd spiral modes, the two bar halves typically do not connect at exactly the same time to a spiral arm, and their individual lengths can be significantly offset. We estimated that in about 50\% of bar measurements in Milky Way-mass external galaxies, the bar lengths of SBab type galaxies are overestimated by $\sim15\%$ and those of SBbc types by $\sim55\%$. Consequently, bars longer than their corotation radius reported in the literature, dubbed ``ultra-fast bars", may simply correspond to the largest biases. Given that the Scutum-Centaurus arm is likely connected to the near half of the Milky Way bar, recent direct measurements may be overestimating its length by $1-1.5$~kpc, while its present pattern speed may be $5-10$~\ksk smaller than its time-averaged value.
\end{abstract}




\section{Introduction}

Galactic bars reside in the centres of about 2/3 of nearby spiral galaxies, as seen in the near-infrared (e.g., \citealt{eskridge00}). Bars are typically described by their length, strength, and pattern speed. Their length can be estimated visually  (e.g., \citealt{martin95}), by structural decompositions of the galaxy surface brightness (e.g., \citealt{dejong96, prieto97, gadotti11}), by locating the maximum in the isophotal ellipticity (e.g., \citealt{wozniak95, laine02, aguerri09}), by variations of the isophotal position angle (e.g., \citealt{sheth03}), or by variations of the Fourier modes phase angle of the galaxy light distribution (e.g., \citealt{quillen94}). 
Bar lengths have been found to correlate with galaxy parameters, such as the galaxy mass, galaxy color, the disc scale-length, and the bulge size (e.g., \citealt{aguerri05, marinova07, gadotti11}). Early-type systems host significantly larger bars than late-type ones (e.g., \citealt{elmegreen85, menemdez07, aguerri09}). 

The bar angular velocity (or pattern speed, $\Omega_\text{b}$ or $\Omega_\text{p}$) determines at what radii resonances occur in the disk, knowledge of which is necessary to understand the bar's impact on the disk dynamics. While bar length and strength can be directly measured from the observations, estimating $\Omega_\text{b}$ in principle requires kinematic information. To get around this, indirect methods have been developed, e.g., by identifying rings in the disk morphology with the location of the Lindblad resonances or sign-reversal of streaming motions across the corotation radius (CR, e.g., \citealt{buta86,  jeong07}). A model-independent direct measurement of $\Omega_\text{b}$ using kinematics is the Tremaine-Weinberg method (\citealt{tremaine84}, hereafter TW). This has been applied extensively to individual external galaxies (e.g., \citealt{merrifield95, aguerri03, meidt09}), SDSS-IV MaNGA IFU data \citep{bundy15, guo19}, the CALIFA survey \citep{sanchez12, aguerri15}, as well as to the Milky Way (hereafter MW, \citealt{debattista02, sanders19b, bovy19}).

Unlike in external galaxies, the MW bar is hard to observe directly owing to our position in the disk plane, therefore, indirect approaches have been used to determine its length, strength, orientation, and pattern speed (for a review, see \citealt{bland-hawthorn16}). Until about five years ago, the bar half-length was thought to be well constrained to $R_{\rm b}\sim3.5$~kpc and its pattern speed to $\Omega_\text{b}\sim50-60$~\ksk, based on matching longitude-velocity ($\ell-V$) diagrams of {H}{I} and CO gas in the inner MW \citep{englmaier99}, the position of the Lagrangian point $L_4$ \citep{binney97}, the position of the Hercules stream in the $u-v$ plane \citep{dehnen00, fux01, antoja12, monari17b}, the Oort constant C \citep{mnq07, siebert11b, bovy15}, and some low-velocity moving groups in the $u-v$ plane \citep{minchev10}, although lower $\Omega_\text{b}$ estimates did exist (e.g., \citealt{weiner99,rodriguez08}).

Starting with \cite{wegg15}, \cite{sormani15}, \cite{li16}, and \cite{portail17}, more recent works using different datasets and methods have suggested a significantly longer bar than previously thought ($\sim5$~kpc) and a pattern speed much lower than previously accepted ($35-45$ \ksk, e.g., \citealt{hunt18, sanders19b, clarke19, monari19, bovy19}). In contrast, \cite{anders19} found a bar-shaped feature inclined by $\sim40^\circ$ with respect to the solar azimuth and a length of $\sim3.5$~kpc in the stellar density distribution of Gaia DR2 data \citep{gaia18} for stars brighter than $G=18$, using distances derived with the StarHorse code \citep{santiago16, queiroz18}. Some studies find consistency with both a slow and a fast bar \citep{hattori19, trick19}.

To some degree, such disparity may result from the different methods used to measure the bar length.
There could be also dynamical reasons for finding different bar lengths and pattern speeds, as we will argue in this work. \cite{quillen11} noted that in their N-body simulations the bar length visibly fluctuates in $R-\phi$ density maps, resulting from the interaction with the inner disk spiral structure, as spirals connect and disconnect from the bar ends. Time dependent fluctuations in bar length, strength, and pattern speed were found in double-barred N-body models by \citet{wu18}, interpreted as the interaction between the two bars moving with different pattern speed. 

The present work studies two hydrodynamical simulations of MW-like disks forming in the cosmological context, in an effort to quantify variations in bar parameters on a dynamical timescale. Implications for the MW and external galaxies are discussed.

This paper is organized as follows. In \S\ref{sec:sims} we describe our two simulations and in \S\ref{sec:barmethods} our three methods of bar-length measurement are introduced. In \S\ref{sec:time} we quantify the time oscillations of the bars' lengths, amplitudes, and pattern speeds. Interpretation for these fluctuations is offered in \S\ref{sec:barspirals}, where we perform power spectrum analyses relating bar oscillation frequencies to the reconnection between bars and spiral modes of different multiplicity. A comprehensive discussion is presented in \S\ref{sec:discussion}, where we relate to other numerical work and make predictions for both observations of external galaxies and the MW. Finally, we conclude with a summary in \S\ref{sec:conclusions}.

\section{Simulations}
\label{sec:sims}

We consider the last $1.38$~Gyr of evolution before redshift zero from two simulations in the cosmological context with disk properties close to those of the MW, e.g., both having central bars, velocity dispersion radial profiles compatible with observations, and the presence of spiral arms. 

The first simulation was first presented by \citet{buck18} and is out of a suite of high-resolution hydrodynamical simulations of MW-sized galaxies from the NIHAO-UHD project (\citealt{buck19c}, galaxy g2.79e12, hereafter Model1). This galaxy was simulated using a modified version of the smoothed particle hydrodynamics (SPH) solver GASOLINE2 \citep{wadsley17} and star formation and feedback are modelled following the prescriptions in \citet{stinson06} and \citet{stinson13a}. The total stellar mass of Model1 is $1.59\times10^{11}~M_\odot$. The galaxy is resolved with $\sim8.2\times10^6$ star, $\sim2.2\times10^6$ gas, and $\sim5.4\times10^6$ dark matter particles \citep[Table 1 in][]{buck19a}, which corresponds to a baryonic mass resolution of $\sim3\times10^4~M_\odot$ per star particle ($\sim9\times10^4~M_\odot$ gas particle mass) or 265~pc force softening. For more details on the simulation details and galaxy properties we refer the reader to \cite{buck19a}.

Model1 was also used to study the chemical bimodality of disk stars \citep{buck20} and its satellite galaxies closely follow the observed satellite mass function \citep{buck19a}.
To properly study the time evolution of the disk's central region we require closely spaced time outputs, here using snapshots every 6.9~Myr. This ensures that the central barred region (where the period is $\sim100$~Myr) would have over a decade of complete rotations.

The second model is from a suite of 33 simulations presented by \cite{martig12} (the g106 galaxy, hereafter Model2) and also studied extensively in the past (e.g. \citealt{martig14a, martig14b}, \citealt{kraljic12}, \citealt{mcm13, mcm14, minchev14, minchev15}, \citealt{carrillo19}). Time outputs here are separated by 4.5~Myr. The simulation is run using a re-simulation technique first introduced in \cite{martig09} and the Particle Mesh code described by \cite{bournaud02, bournaud03}. The spatial resolution is 150~pc and the mass resolution is $3\times10^5~M_\odot$ for dark matter particles, $7.5\times10^4~M_\odot$ for star particles present in the initial conditions, and $1.4\times10^4~M_\odot$ for gas particles and star particles formed during the simulation. Model2 has a stellar mass of $\sim4.3\times10^{10}~M_\odot$ (within the optical radius of 25~kpc) and a dark matter mass of $\sim3.4\times10^{11}~M_\odot$.

Originally Model1 and Model2 have disk scale-lengths of $h_d\approx5.6$ and $h_d\approx5.1$~kpc and roughly flat rotation curves at $V_{\rm c}\approx340$ and $V_{\rm c}\approx210$~\kms, respectively. We rescaled both models' positions and velocities in order to match measurements for the MW: $h_d=3.5$~kpc and $V_{\rm c}=240$~\kms \citep{bland-hawthorn16}, which affects the mass, $M$, of each particle according to the relation $GM\sim V^2R$, where $G$ is the gravitational constant. We chose an $h_d$ value near the upper limit of the recommendation by \cite{bland-hawthorn16} so that the bar lengths do not become too short.

\begin{figure*}
	\includegraphics[width=1.8\columnwidth]{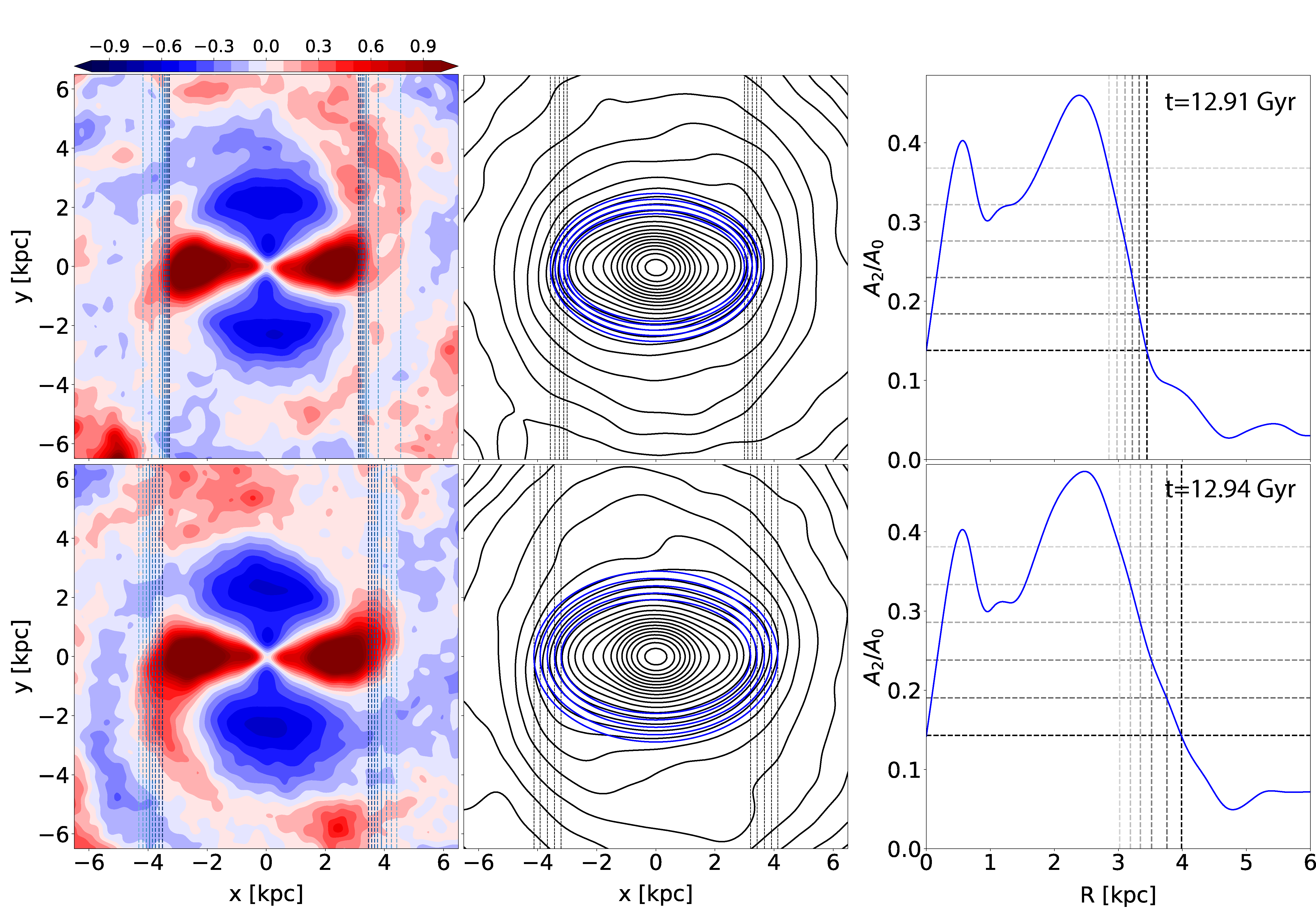}
    \caption{Illustration of the three methods used to measure the bar length. For the top row we use a snapshot from Model1 at $t=12.91$~Gyr and for the bottom one an output $\sim35$~Myr later. {\bf Left column:} $L_{\mathrm{cont}}$ method. Face-on stellar density contours with the azimuthally-averaged density subtracted. Vertical dashed lines mark the contour levels crossing the y-axis where the overdensity of stars has dropped by 10\% to 80$\%$ from the maximum. 
    {\bf The 50\% drop is shown by the solid vertical lines. }
    {\bf Middle column:} $L_{\mathrm{prof}}$ method. Bar length is measured by fitting ellipses and measuring when the difference between the density along the semi-major and semi-minor axis falls to below 30\% to 40\% of the semi-major value. {\bf Right column:} $L_{m\mathrm{=2}}$ method uses the ratio of the amplitude of the $m=2$ to the $m=0$ Fourier components of the stellar density, $A_2/A_0$, as a function of radius, $R$. The bar length is taken as the radius where this ratio falls below a fraction of the maximum value. We here consider six thresholds from 30\% to 80$\%$ of the maximum value. The larger bar estimate in the bottom row is due to its being connected to spiral arms, as seen in the left column.
    }
    \label{fig:Methods}
\end{figure*}

\subsection{Bars}

After the rescaling, both Model1 and Model2 have very similar bar lengths\footnote{Estimated from the $L_{\rm cont}$ method, described in \S\ref{sec:lcont}.} at the final time, $R_{\rm b}\approx3.05$~kpc and $R_{\rm b}\approx3.2$~kpc, respectively, but arrive there by different paths. During the period studied, Model1's bar length decreases monotonically by about 10\% while that of Model2 increases by the same amount (see dotted-red lines in Figs.~\ref{fig:Lcont} and \ref{fig:Lcont-all-Martig}). As may be expected, the pattern speeds change in the opposite directions with final values of $\sim80$ and $\sim50$~\ksk, respectively. The bar lengths quoted above are the ``true" values, the meaning of which will become clear in the next sections. Typically bars are found to slow down and to grow in length with time (as in Model2) due to losing angular momentum to the disk and dark mater halo. The opposite behavior of Model1's bar is due to gas infall at this particular time period, given that the simulation is unconstrained and in the cosmological context.

Having the same length but very different pattern speeds places the bar resonances at very different radii for each simulation. This makes Model1 comparable to the fastest bars found in observations, given by the ratio of the bar's CR radius to its length, $\mathscr{R}\equiv R_{CR}/R_{\rm b}\approx3.1/3.05\approx1.02$; conversely, Model2 hosts a significantly slower bar, with $\mathscr{R}=5.6/3.2\approx1.75$ (e.g., see Table~1 by \citealt{rautiainen08}), using final time values.

\subsection{Spiral structure}
\label{sec:sp}

The spirals of Model1 are more tightly wound and multi-armed (see Fig.~1 in \citealt{buck18}), while for Model2 they are more open and dominated by two or four arms (see top-right panel of Fig.~1 in \citealt{martig14a} or Fig.~1 in \citealt{mcm13}), which signifies that they are stronger. Indeed, we measured spiral structure overdensity for Model1 typically $\sim5-10\%$ higher than the background, compared to $\sim15-25\%$ for Model2 (see rightmost columns of Figs.~\ref{fig:Methods} and \ref{fig:Methods2}). These values are on the lower end of the 15\%-60\% spiral-arm overdensity estimated by \cite{rix95} for 18 face-on spiral galaxies. 

Recent estimates of the MW spiral-arm overdensity include $\sim14\%$ from modeling the radial velocity field of RAVE data \citep{siebert12}, $\sim26\%$ needed to account for the migration rate of supersolar metallicity open clusters near the Sun \citep{quillen18a}, and $\sim20\%$ obtained from matching the radial velocity field of stars on the upper red giant branch from a compilation of data \citep{eilers20}. These are somewhat larger than the spiral strength of our Model1 and quite consistent with our Model2.

\section{Measurements of bar length}
\label{sec:barmethods}

Here we employ three methods to determine the bar length\footnote{Hereafter we use ``bar length" to mean the length of its semi-major axis, as frequently done in the literature.} of Model1, two of which have been widely used in the literature (e.g., \citealt{athanassoula02, wegg15, wu18}) and a new approach introduced below. We use a cut of $|z|<1$~kpc, where $z$ is the distance from the disk midplane, but the results do not vary wildly for other reasonable values.

\subsection{Drop in background-subtracted densities: $L_{\rm cont}$}
\label{sec:lcont}

Since galactic bars feature very high stellar densities relative to the rest of the disk, the surface density along the bar major axis will start to drop approaching the bar ends, as can be seen in the left column of Fig.~\ref{fig:Methods}, which shows the background-subtracted surface density plots for two time outputs from Model1 35~Myr apart. The density will then either fall off until it matches that of the disk (the case for the top panel) or it would be elevated if spiral structure is present nearby (as in the bottom panel). 

Our new method of measuring the bar length is somewhat similar to tracing the drop of the $A_2/A_0$ Fourier component (see \S\ref{sec:lm}), but uses instead a drop in the background-subtracted density, considering the range between 10\% and 80$\%$ above the radial mean. This range, covered in steps of 10\%, is shown by the vertical dashed lines, indicating the corresponding contour levels.

An important feature of the method is that it allows one to estimate each side of the bar separately. From the left column of Fig.~\ref{fig:Methods} it is already obvious that only in a time range of 35~Myr the bar can change length by about 10-20\%, which varies depending on the choice of threshold.

\begin{figure*}
	\centering
	\includegraphics[width=1.8\columnwidth]{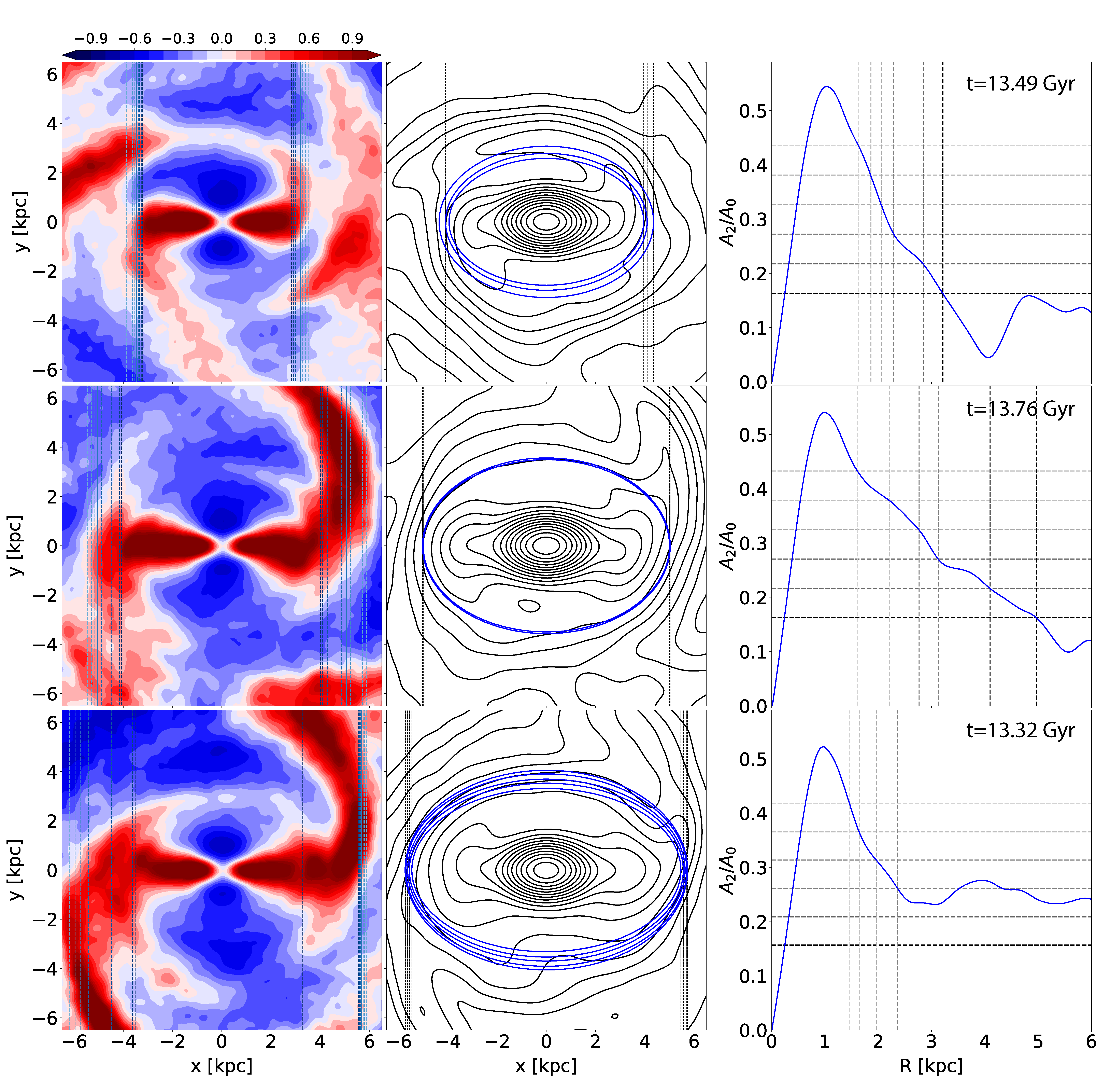}
    \caption[]{As Fig.~\ref{fig:Methods}, but for Model2. We have chosen three snapshots to highlight some typical cases. 
    {\bf Top row:} At $t=13.49$~Myr, the bar is relatively well separated from spiral arms, as seen in the background-subtracted density, and the $L_{\rm cont}$ measured length lies in the range $\sim2.8-3.5$~kpc, depending on the threshold used. This range changes to $\sim4-4.2$~kpc for $L_{\rm prof}$ and $\sim1.6-3.2$~kpc for $L_{\rm m=2}$, i.e., the three methods are much less consistent than for Model1. We tend to trust the $L_{\rm cont}$ method more than the other two, since we can clearly see the bar morphology and orientation with the spiral structure for each  bar half.
{\bf Middle row:} At a time output $\sim270$~Myr later, the bar is measured to be 25-50\% longer, with artifacts due to connected spiral arms clearly seen in the left panel, but not in in the total density contours in the middle or the azimuthally averaged $A_2/A_0$ variation with radius in the right panel. 
{\bf Bottom row:} At a time output $\sim170$~Myr earlier than in the top row, the bar is overestimated by almost a factor of two by $L_{\rm cont}$ and by $\sim38\%$ by $L_{\rm prof}$. This problem is evident from the overdensity discontinuity in the right bar half of the $L_{\rm cont}$ measurement, but is not clear from the other two methods. In both the second and third rows the bar is connected to the spiral structure. As for Model1, we take the $L_{\rm cont}$ measurement with threshold of 50\% to be the ``true" bar length at $t=13.49$~Gyr, $R\approx3.1$~kpc.
    }
    \label{fig:Methods2}
\end{figure*}

\subsection{Drop in disk ellipticities: $L_{\rm prof}$}
\label{sec:lprof}

The density profiles along the bar major and minor axes gradually become similar as radius increases. \citet{athanassoula02} proposed to fit ellipses to the central density region while gradually increasing radius until reaching a point where the density along the semi-minor and semi-major axes are the same within 5\%. 

This method is adapted here, though with the threshold modified to use a range of ellipticities between 30$\%$ and 40$\%$ in steps of 2.5$\%$ of the difference between the bar major and minor axes. This range covers the value used by \citet{wegg15} (30$\%$) to estimate the length of the MW bar in their N-body model, which was significantly higher than the 5\% of \citet{athanassoula02}. We also agree with a larger threshold, as we found that a smaller one often produced abnormally large values or failed altogether. 

We show the results of this bar length measurement in the middle column in Fig.~\ref{fig:Methods} over the stellar density contours.
As in the left column, the $L_{\rm prof}$ method measures a longer bar for the bottom row time output by about a similar amount.

\subsection{Fourier analysis of the central disk: $L_{\rm m=2}$}
\label{sec:lm}

An estimate of bar length can also be obtained by taking the Fourier transform over all disk azimuths. This can find the numbers, strengths, and multiplicities of non-axisymmetric modes \citep{masset97, meidt08, quillen11}. For each disk component being analysed, the following coefficients of the Fourier series are first determined:

\begin{equation}
\begin{split}
    a_{m}(R) = \frac{1}{\pi}\int^{2\pi}_{0}\rho(R,\theta)~ cos(m\theta)~d\theta\\
    b_{m}(R) = \frac{1}{\pi}\int^{2\pi}_{0}\rho(R,\theta)~ sin(m\theta)~d\theta
\end{split}
	\label{eqn:fourier}
\end{equation}

Here, $m$ is the azimuthal wavenumber and $\rho(R,\theta)$ is the mass density at a specific spatial bin. We estimate $A_m$ with respect to the axisymmetric component $A_0$, as $\sqrt{a_m^2+b_m^2}/A_0$ (see, e.g., \citealt{athanassoula02}).

Any galactic bar, which would have rotational symmetry of order two will, therefore, be highlighted in the $m=2$ Fourier component, along with any 2-armed spiral structure. This allows the bar strength to be seen as a function of radius in the rightmost column in Fig.~\ref{fig:Methods}, using 300~pc radial and $10^{\circ}$ azimuthal bins. The bar length is estimated from the radius at which $A_2/A_0$ drops below some percentage of the maximum strength in the range 30\% to 80\%, in steps of 10\%. Measuring bar lengths of individual sides could then be done by reflecting one half of the disk onto the other side prior to performing the analyses. As in the previous two methods, it is clear from the right column of Fig.~\ref{fig:Methods} that the bar length measured at the second time output is longer by 10\%-20\%, depending on the threshold used. 

\begin{figure*}
	\centering
	\includegraphics[width=1.8\columnwidth]{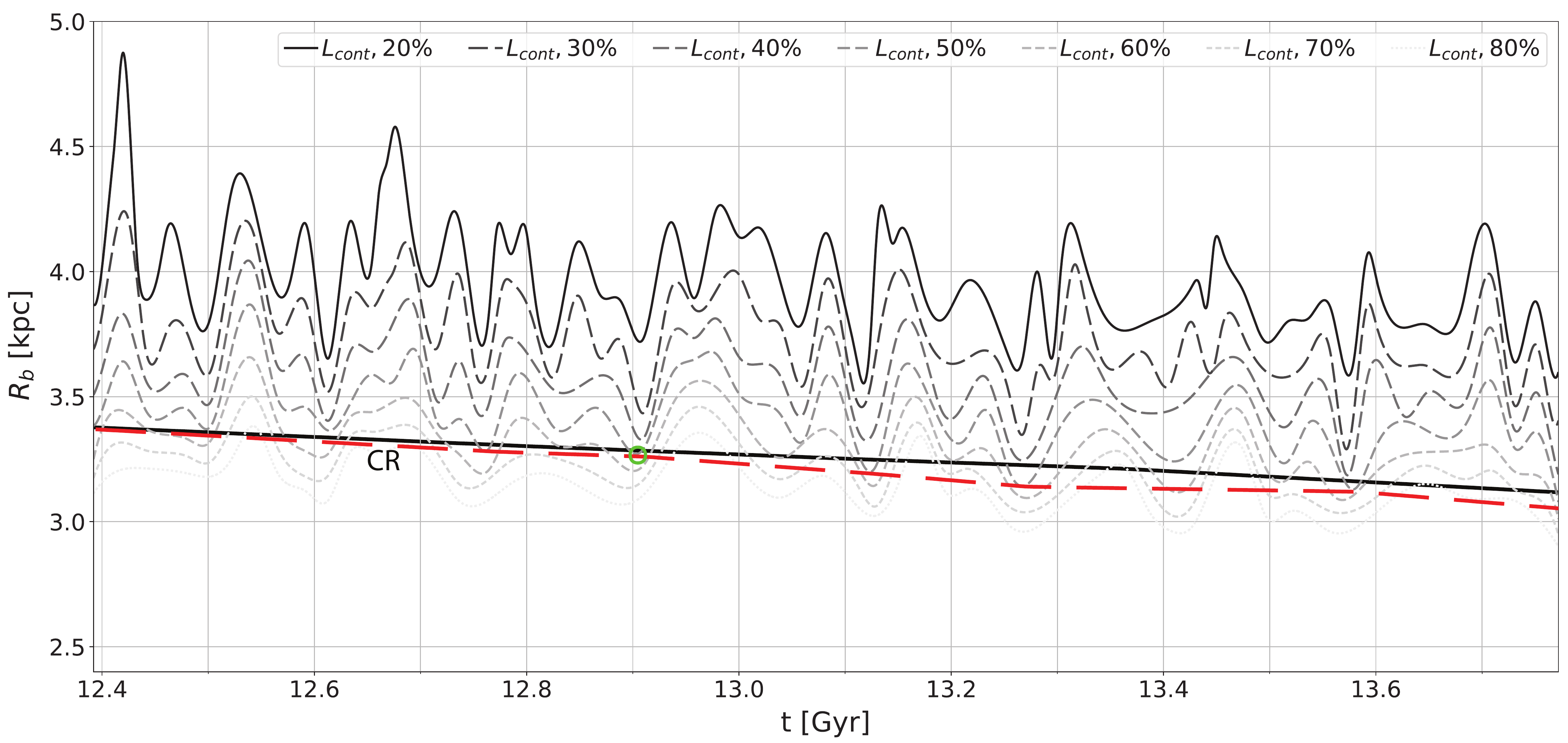}
    \caption[]{Variations with time of the $L_{\rm cont}$ bar length measurement for Model1 (see Fig.~\ref{fig:Methods}). Light smoothing is applied. Different curves show different threshold values in density between 20\% and 80\%, as indicated on top. Smooth variations with a period $T_{long}\approx125$~Myr are seen for the smaller threshold values, however, additional peaks appear for the largest three thresholds with $T_{short}\approx T_{long}/2$. The green circle marks the time output used in the top row of Fig.~\ref{fig:Methods}, which corresponds to a local minimum. Upon inspection of all measurement methods in Fig.~\ref{fig:Methods}, we choose the $L_{\rm cont}$ threshold of 50\% to refer to as the ``true" bar length at this time ($R_{\rm b}\approx3.25$~kpc). 
    To get the true bar length as a function of time we interpolate over the minimum values for the same threshold (red-dashed line), corresponding to when the bar and spirals are not connected. 
       The bar is clearly seen to get shorter with time, starting with $R_{\rm b}\approx3.35$~kpc and ending up with $R_{\rm b}\approx3.05$~kpc in the period of time we consider. The decrease in bar length with time is accompanied by an increase in pattern speed (see Fig.~\ref{fig:tw2}), such that the CR radius follows closely the bar's length. To see this, we overlaid the evolution of the mean CR radius (solid black curve marked by ``CR" ), estimated form the $m=2$ Fourier component in power spectrograms (see \S\ref{sec:power}).
    }
    \label{fig:Lcont}
\end{figure*}

For our choice of thresholds, the three methods agree quite well in the ranges of bar length they measure for Model1, however, this is not going to be the case for Model2. As will become clear from Fig.~\ref{fig:Lcont}, the top row of Fig.~\ref{fig:Methods} shows a time when the bar length is at a local minimum. We can argue that this then represents the ``true" bar length, while the larger bar measurement in the bottom row of Fig.~\ref{fig:Methods} is caused by connecting spiral arms (this is much more obvious for Model2, see \S\ref{sec:mod2len}). We, therefore, take the ``true" bar length at $t=12.91$~Gyr to be $L_{\rm cont}$ with threshold of 50\%, which corresponds to $R_{\rm b}\approx3.25$~kpc. Note that this changes monotonically with time owing to the bar's secular evolution, as seen in Fig.~\ref{fig:Lcont}.

\subsection{Model2 bar length measurements}
\label{sec:mod2len}

In Fig.~\ref{fig:Methods2} we present the three bar length measurements applied to Model2, as done for Model1 in Fig.~\ref{fig:Methods}. We have chosen three snapshots to highlight some typical cases, since this galaxy shows more complex variations than Model1. In the top row, $L_{\rm cont}$ shows that the bar is relatively well separated from spiral arms and the measured length is 2.8-3.5~kpc, depending on the drop. The left bar half is about 10\% longer owing to a connected spiral, as evident from the disturbed highest density contour.

\begin{figure*}
	\centering
	\includegraphics[width=1.8\columnwidth]{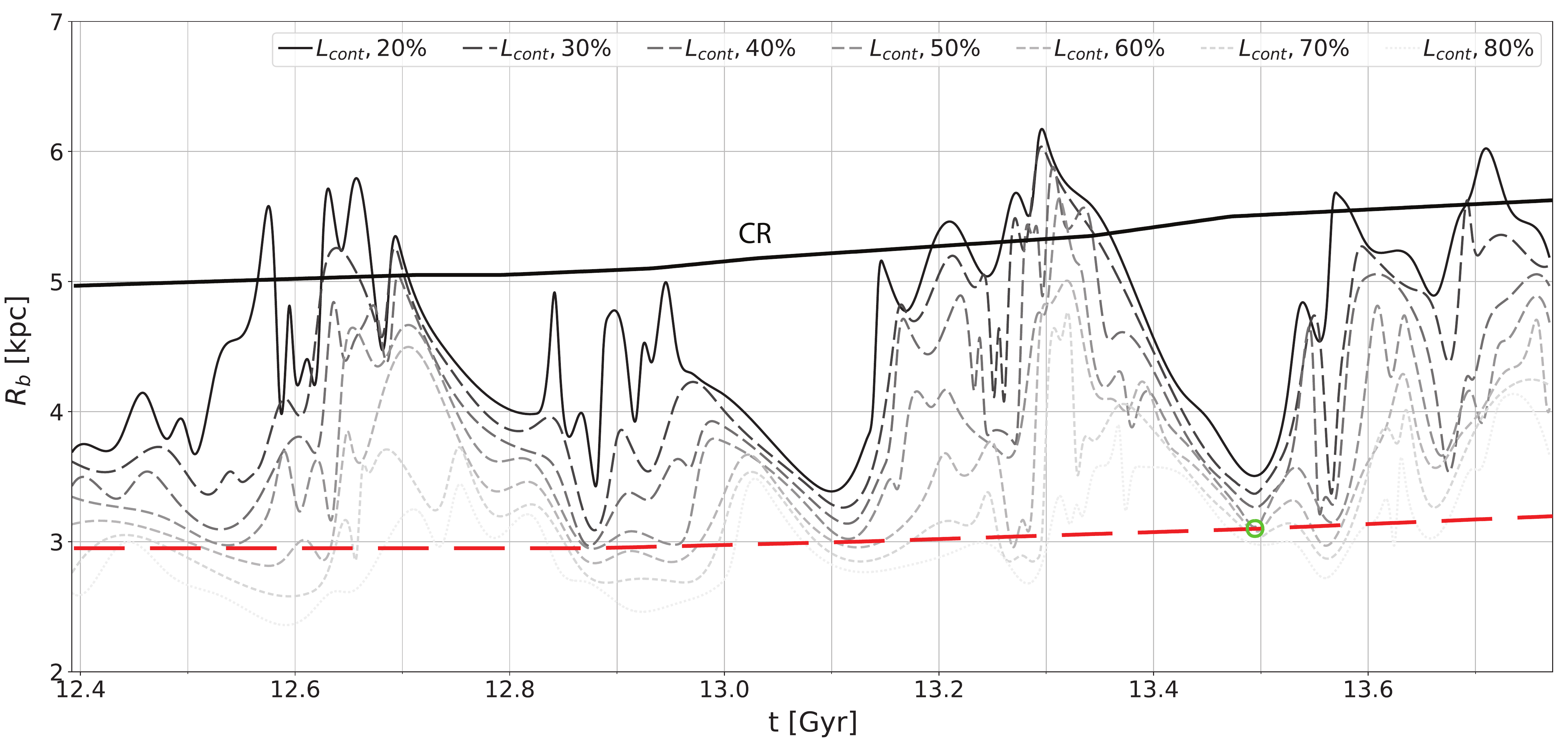}
    \caption[]{
    As Fig.~\ref{fig:Lcont}, but for Model2. The $L_{\rm cont}$ bar length measurement varies more erratically with time compared to Model1, as already expected from Fig.~\ref{fig:Methods2}. The period is also less regular than for Model1 and we find a longer period overall, due to the slower bar here. 
    The green circle marks the time output used in the top row of Fig.~\ref{fig:Methods2}, which corresponds to a local minimum. As for Model1, we use $L_{\rm cont}$ with a threshold of 50\% to get the ``true" bar length at this time, finding $R_{\rm b}\approx3.1$~kpc. The red-dashed line results from interpolating over such minima, which then gives the secular evolution of the true bar length. Opposite to Model1, the bar size increases with time from $R_{\rm b}\approx2.9$~kpc to $R_{\rm b}\approx3.2$~kpc in the period of time we consider.
    The time evolution of the CR radius is shown by the solid-black line, found to be at a much larger radius than the bar length, compared to Model1. This is because of the lower bar pattern speed for Model2 (see Figs.~\ref{fig:tw2} and \ref{fig:power2}).
    }
    \label{fig:Lcont-all-Martig}
\end{figure*}

The $L_{\rm prof}$ measurement is systematically larger, around 4 kpc. Conversely, $L_{\rm m=2}$ varies between about 1.6 and 3.2 kpc. Note the strong disagreement among the different bar estimates compared to Model1, although the same thresholds were used for each model. We conclude that the ``true" bar length is given by the bar side along the positive $x-$axis of $L_{\rm cont}$, as the disturbance seen in the left half must be caused by a spiral arm. As for Model1, we take the ``true" bar length at this time to be the $L_{\rm cont}$ value with threshold of 50\%, which corresponds to $R_{\rm b}\approx3.1$~kpc. We note that this varies monotonically with time due to the bar's secular evolution, which can be seen in Fig.~\ref{fig:Lcont-all-Martig}.

The second row of Fig.~\ref{fig:Methods2} shows a case $\sim270$~Myr later, where the bar is measured to be 25\%-50\% longer by all three methods. Finally, the third row shows a time output $\sim170$~Myr earlier than in the top row, where the bar is found to be larger by almost a factor of two with the $L_{\rm cont}$ and by $\sim38\%$ with $L_{\rm prof}$ methods. For $L_{\rm m=2}$ the $m=2$ amplitude does not drop below 50\% for the radial range of 6~kpc shown in the plot. This means that this method would measure a length $>6$~kpc for larger drops in $A_2/A_0$, which would clearly be incorrect. Comparing $A_2/A_0$ between the two models reveals also that Model2 has a significantly stronger spiral arm overdensity. 

From the $L_{\rm cont}$ plot in the bottom row of Fig.~\ref{fig:Methods2} it is obvious that the spiral arm orientation is such that it adds to the bar length, however, in the case of $L_{\rm prof}$ even a visual inspection would not catch this problem, since the density variations along the bar major axis are not seen in the total density, shown as the contours in the middle column. \cite{athanassoula02} warned about using the latter method blindly, as in certain cases there may not be a steep drop. Here, however, we do see a steep drop for the $L_{\rm prof}$ method (in the middle row the 5 thresholds are on top of each other), yet from the $L_{\rm cont}$ measurement we clearly see that the Model2 bar is not longer than $\sim3$~kpc. We attribute this discrepancy to the stronger spiral arms in this hydrodynamical simulation as opposed to the dissipationless N-body runs by \cite{athanassoula02} and other works.
 
Looking only at the $L_{\rm prof}$ measurement (middle column of Fig.~\ref{fig:Methods2}) one could conclude that the bar is much larger in the bottom panel, however, both the $L_{\rm cont}$ and $L_{\rm m=2}$ plots argue against this. It is clear from the $L_{\rm cont}$ plot that the extension of the bar, especially of its right side, comes from the strong spiral arms attached to it. Using only the $L_{\rm prof}$ method can, therefore, lead to erroneous results, as a randomly selected time output may correspond to a time when the bar and spiral structure are connected. Note that when spiral arms are stronger (as in our Model2), $L_{\rm prof}$ tends to overestimate the bar size even when the bar is as best as possible separated from spiral arms, as in the top row of Fig.~\ref{fig:Methods2}. This conclusion is in agreement with the results of \cite{petersen19}, who showed that the bar length measured from the extent of the $x_1$ orbits (true length) is at or below the minimum of their ellipse-fit-derived length (similar to our $L_{\rm prof}$) in their N-body simulations (see their Fig.~10).

\section{Time oscillations of bar parameters}
\label{sec:time}

\subsection{Mean bar length}
\label{sec:mean}

In Fig.~ \ref{fig:Lcont} we plot the bar length time evolution, $R_{\rm b}(t)$, for Model1, using the $L_{\rm cont}$ method described in \S\ref{sec:barmethods}. Variations with a well-defined period are seen over these last $1.38$~Gyr of quiescent disk evolution, which is also true for the other two methods (see Figs.~\ref{fig:Lprof} and \ref{fig:Lm}). The amplitude is typically 0.3~kpc (or $\sim10\%$), decreasing (increasing) for thresholds that measure a shorter (longer) bar for all three methods. Note however that, as discussed below, the deviations from the ``true" bar length are double that, or $\sim20\%$, since the ``true" bar length is given by the minimum of the time variations.

We see in Fig. ~\ref{fig:Lcont} that the time variations for smaller density drops have a period of $T_{long}\approx125$~Myr (gray short-dashed curves), from counting 11 peaks in the period of $1.38$~Gyr. As the drop increases to $50\%$-40\% and beyond, a doubling in the frequency is seen resulting in a period of $T_{short}\sim T_{long}/2$. The more frequent oscillations appear as we enter the disk and encounter different spiral modes of various multiplicity (as will be detailed in \S\ref{sec:power} below). These are not seen from the $L_{\rm prof}$ measurement in Fig.~\ref{fig:Lprof}, except possibly for the 30\% threshold, which may be because fitting an ellipse averages strongly over the density variation. The $L_{\rm m=2}$ method (Fig.~\ref{fig:Lm}), on the other hand, matches quite well the $L_{\rm cont}$ variations with time, including the transition from the lower to higher frequency with increasing density drop, for the most part.

To get the ``true" bar length as a function of time we interpolate over the minimum values measured by the $L_{\rm cont}$ method (corresponding to when the bar and spirals are not connected). The small green circle in Fig.~\ref{fig:Lcont} marks the time output used in the top row of Fig.~\ref{fig:Methods}, which corresponds to a local minimum. Upon inspection of all measurement methods in Fig.~\ref{fig:Methods}, we choose the $L_{\rm cont}$ threshold of 50\% to refer to as the ``true" bar length at this time ($R_{\rm b}\approx3.25$~kpc). The red-dashed line shows the approximate position of the minima for this threshold at different times, which we argued in \S\ref{sec:mod2len} correspond to the true bar length. The bar is seen to decrease with time from $R_{\rm b}\approx3.35$~kpc to $R_{\rm b}\approx3.05$~kpc in the period of time we consider (red-dashed line in Fig.~\ref{fig:Lcont}). The monotonic change is accompanied by an increase in pattern speed (see Fig.~\ref{fig:tw2}), such that the bar CR radius follows closely its length. To see this, we overlaid the evolution of the mean CR radius (solid black curve marked by ``CR" ), estimated form the $m=2$ Fourier component in power spectrograms (see \S\ref{sec:power}).

In the case of Model2, the measured bar extent (Figs.~\ref{fig:Lcont-all-Martig} and \ref{fig:Lprof-all-Martig}) appears to vary much more with time than for Model1, although the true bar length is shorter for most of the time - compare dotted-red lines in Figs.~\ref{fig:Lcont} and \ref{fig:Lcont-all-Martig}. The $L_{\rm cont}$ bar length measurement shown in Fig.~\ref{fig:Lcont-all-Martig} varies more erratically with time compared to Model1, as already expected from Fig.~\ref{fig:Methods2}. The period is also less regular than for Model1 and longer overall, because of the slower bar.
The small green circle marks the time output used in the top row of Fig.~\ref{fig:Methods2}, which corresponds to a local minimum. As for Model1, we use $L_{\rm cont}$ with a threshold of 50\% to get the ``true" bar length at this time, finding $R_{\rm b}\approx3.1$~kpc. The red-dashed line results from interpolating over such minima, which then gives the secular evolution of the true bar length. Opposite to Model1, this bar grows larger with time from $R_{\rm b}\approx2.9$~kpc to $R_{\rm b}\approx3.2$~kpc in the period of time we consider. The time evolution of the CR radius is shown by the solid-black line and is found to be at a much larger radius than the bar length, because of the bar's low pattern speed, compared to Model1 (see Figs.~\ref{fig:tw2} and \ref{fig:power2}).

The amplitude of oscillations in Fig.~\ref{fig:Lcont-all-Martig} is up to $1.5$~kpc, which corresponds to $\sim100\%$ overestimation, since the bar's true length is given by the minima. Fig.~\ref{fig:Lprof-all-Martig} shows that $L_{\rm prof}$ overestimates the bar length for all thresholds, while we found that the different methods agreed well for Model1. Unlike in Model1, using larger thresholds does not result in significantly shorter bar estimates, especially near the minima, which are $\sim1$~kpc above the true bar length determined in Fig.~\ref{fig:Lcont-all-Martig}. We attribute this discrepancy to the stronger spiral structure of Model2.

\begin{figure*}
	\includegraphics[width=1.8\columnwidth]{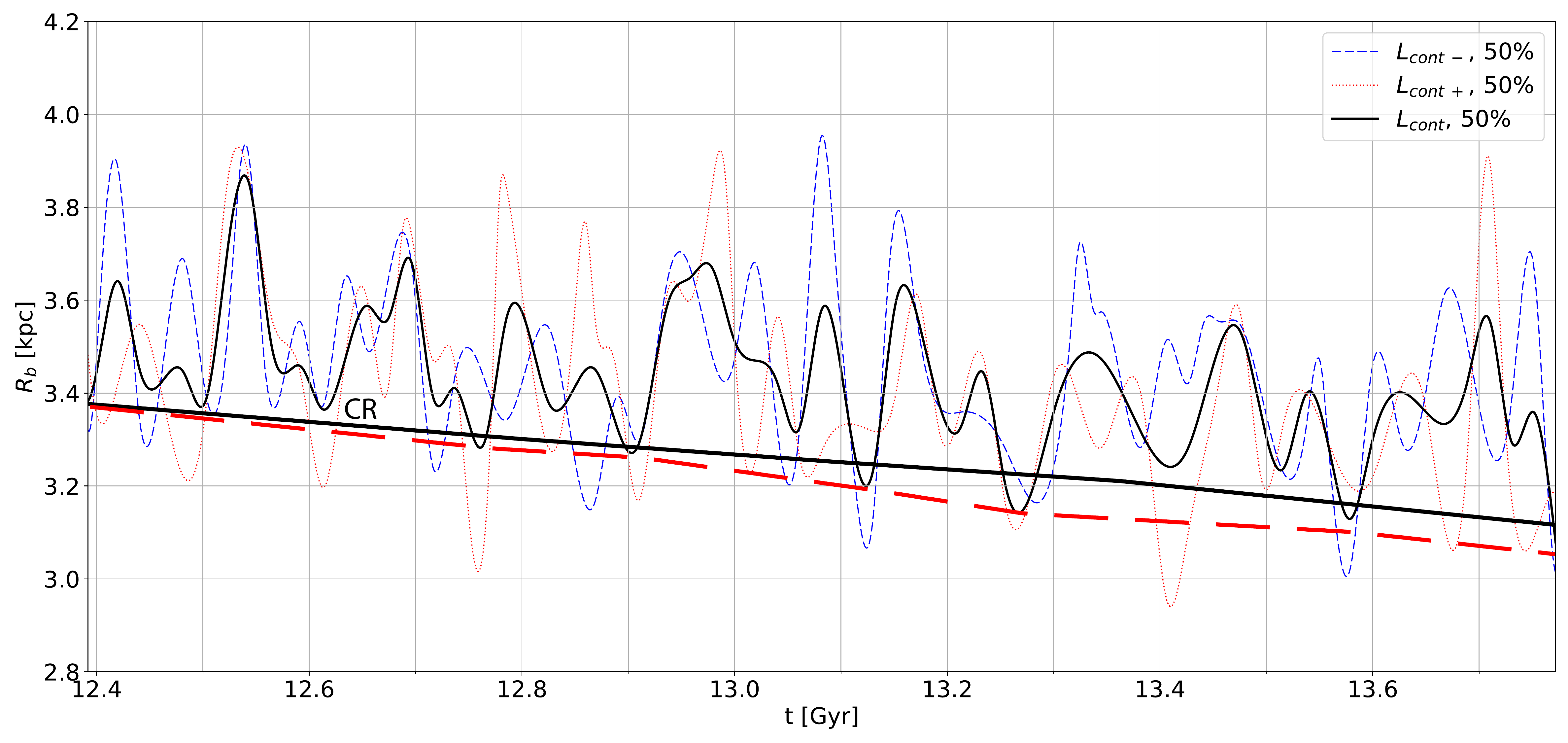}
    \caption{Time variations in individual bar-half lengths for Model1, using the $L_{\rm cont}$ method. $L_{cont-}$ and $L_{cont+}$ correspond to the left and right bar halves, respectively, as seen in Fig.~\ref{fig:Methods}. A density threshold of 50\% is used. The mean bar half-length variation with time is shown by the black curve. Significantly larger fluctuations are found for the individual halves. Individual bar halves peak in length at different times, alternating between smaller and larger maxima. We relate this to the spiral structure's departure from bisymmetry, i.e., the bar ends do not connect to a spiral arm at the same time. The mean CR radius is shown by the black line, marked by ``CR', while the red-dashed line indicates the ``true" bar length.
    }
    \label{fig:Lcont2}
\end{figure*}

\begin{figure*}
	\includegraphics[width=1.8\columnwidth]{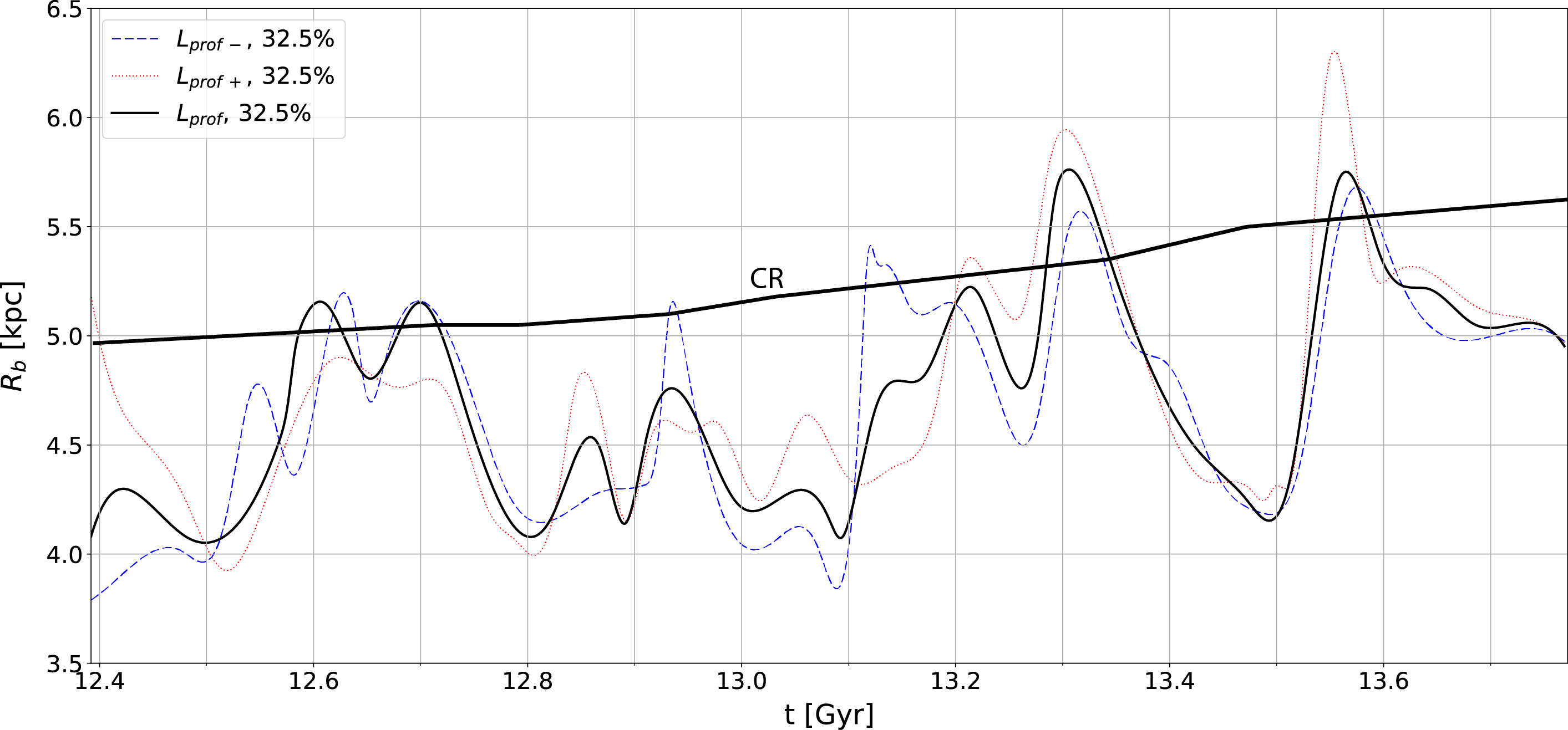}
    \caption{Model2 bar half-length variation with time for the mean and individual ends, using the $L_{\rm prof}$ method. The blue-dashed and red-dotted lines show the left and right halves, respectively, for a bar fixed along the $x-$axis as in Fig.~\ref{fig:Methods2}. Note that the true bar length is $\sim3-3.2$~kpc (see Fig.~\ref{fig:Lcont-all-Martig}), thus, it lies outside the range of this figure. A variation of more than 40\% is seen for this method and threshold, however, the overestimate from the true bar length is $\sim90\%$, considering $R_{\rm b}\approx5.7$~kpc found around 13~Gyr. We argue that this results from the bar-spiral structure overlap at the bar ends. The mean bar CR radius estimated from the $m=2$ Fourier component of power spectrograms (see \S\ref{sec:power}) is shown by the black line, marked with ``CR'. Note that the bar's instantaneous pattern speed fluctuates between $\sim40$ and $\sim65$~\ksk (see Fig.~\ref{fig:tw2}), resulting in CR radius fluctuations in the range $\sim3.8<R_{CR}< 6.4$~kpc. It is remarkable that even this very slow bar can appear ``ultra-fast" for a small fraction of the time.
    }
    \label{fig:Lprof-Martig}
\end{figure*}

\begin{figure*}
	\includegraphics[width=1.8\columnwidth]{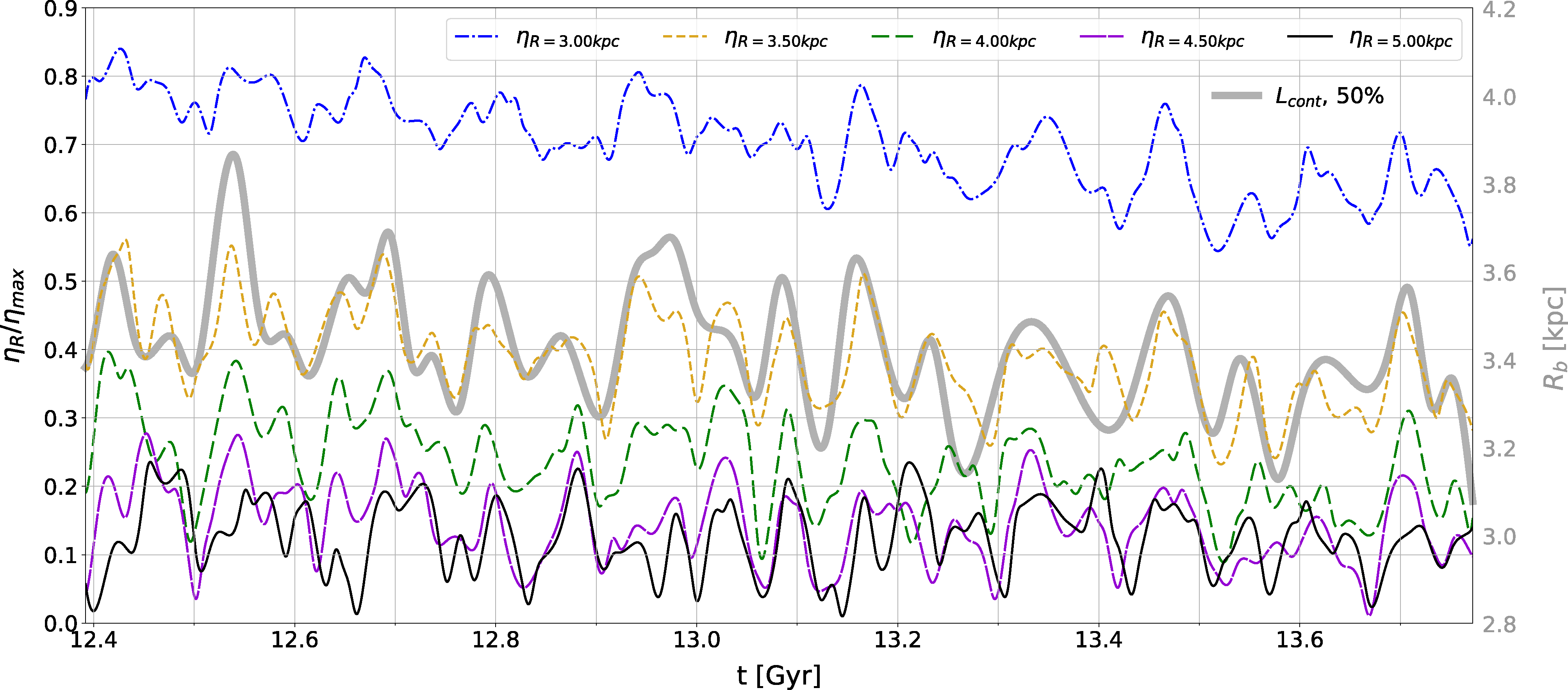}
    \caption{Bar amplitude for Model1 at five different distances from the disk center along the bar major axis, divided by the maximum, $\eta_R/\eta_{max}$, as indicated. Very similar time variations appear for all radii, with typical amplitude changes in $\eta_R/\eta_{max}$ of $\sim0.2$, except for the innermost radial bin. The thick solid-gray curve indicates the mean bar half-length variations (black curve from Fig.~\ref{fig:Lcont2}), showing a very good agreement with the amplitude fluctuations. Note that as the bar decreases in length, so does its amplitude.
    }
    \label{fig:FMax-ends}
\end{figure*}

\begin{figure*}
	\includegraphics[width=1.8\columnwidth]{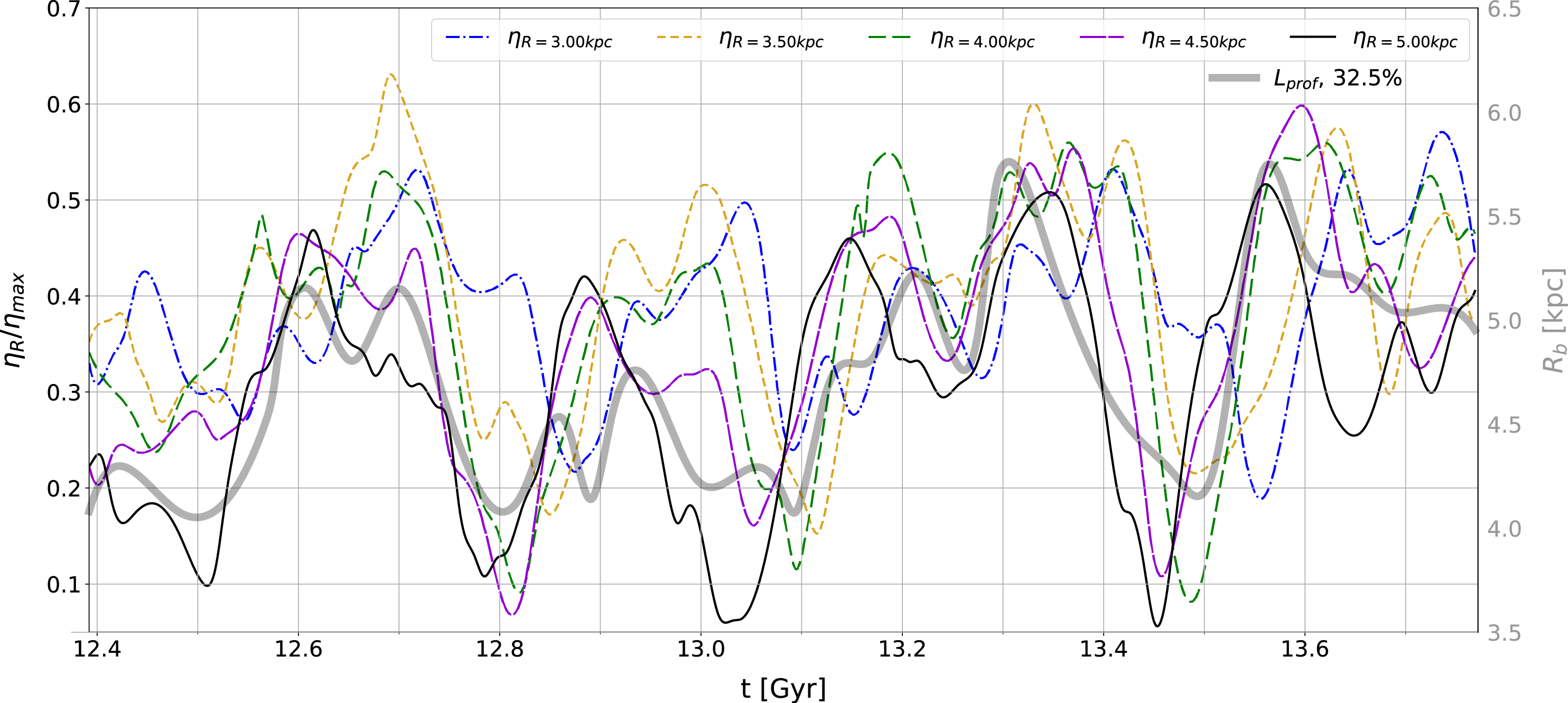}
    \caption{As Fig.~\ref{fig:FMax-ends}, but for Model2. The solid-gray thick curve indicates the mean bar half-length variations (black curve from Fig.~\ref{fig:Lprof-Martig}), showing an overall good agreement with the amplitude fluctuations. Since the true bar length is $\sim3$~kpc, the outermost three radial bins for which $\eta_R$ was estimated correspond to the spiral arms, as evident from the systematic time offset among different curves. 
    }
    \label{fig:FMax-ends-Marig}
\end{figure*}

\subsection{Individual bar halves}
\label{sec:barsides}

Since we would like to link our results to the MW bar, for which our current knowledge is limited to its near end, we also examined each bar half separately. This could be achieved naturally from the $L_{\rm cont}$ method. For the other two methods we reflected the disk density containing the bar side under consideration along the bar minor axis (i.e., in the case of $L_{\rm prof}$, across the line $x=0$ in the middle row of Fig.~\ref{fig:Methods}), after which we applied the method as before. To make sure we did not measure different lengths for different bar halves just because our disk was not centered correctly we did a number of tests. The disk was centered by subtracting the centroid of a cylinder of radius $r_c$ and height $z_c$. We experimented with $r_c$ and $z_c$ values ranging from 2 to 6~kpc and from 0.1 to 1~kpc, respectively, finding that our results were minimally affected.

Figs.~\ref{fig:Lcont2}, \ref{fig:Lprof2}, and \ref{fig:Lm-30} show the time variations in individual bar halves (blue-dashed and red-dotted curves) for Model1 for each of the three measurement methods. For all three methods individual bar halves have larger length fluctuations (sometimes by a factor of two) than the mean bar length variations shown by the black curves. The length can be seen to vary by  $\sim1$~kpc for Model1. As expected, a peak in the length of one side of the bar does not necessarily correspond to a peak in the other - these are often completely offset, i.e., a maximum length measured for one side corresponds to a minimum for the other (e.g., at $t\approx12.78$~Gyr in Fig.~\ref{fig:Lcont2}). It should be noted that the mean bar half-length is not the mean of the individual sides estimated here, except for the $L_{\rm cont}$ method. The other two methods measure the total bar length as described in \S\ref{sec:lprof} and \S\ref{sec:lm} and then divide in half.

Fig.~\ref{fig:Lprof-Martig} shows the variation of individual bar halves for Model2. Here we used the $L_{\rm prof}$ method, since $L_{\rm cont}$ shows abrupt changed on a short timescale (as seen in Fig.~\ref{fig:Lcont-all-Martig}). Similarly to Model1, there is no apparent correlation between the two half-length fluctuations (see \ref{fig:Lprof2}) and often a peak in the length of one bar-half corresponds to a minimum for the other. Larger maxima and minima are reached than those seen in the mean bar length measurement, but not to the extent found for Model1.

\begin{figure*}
	\centering
	\includegraphics[width=2.\columnwidth]{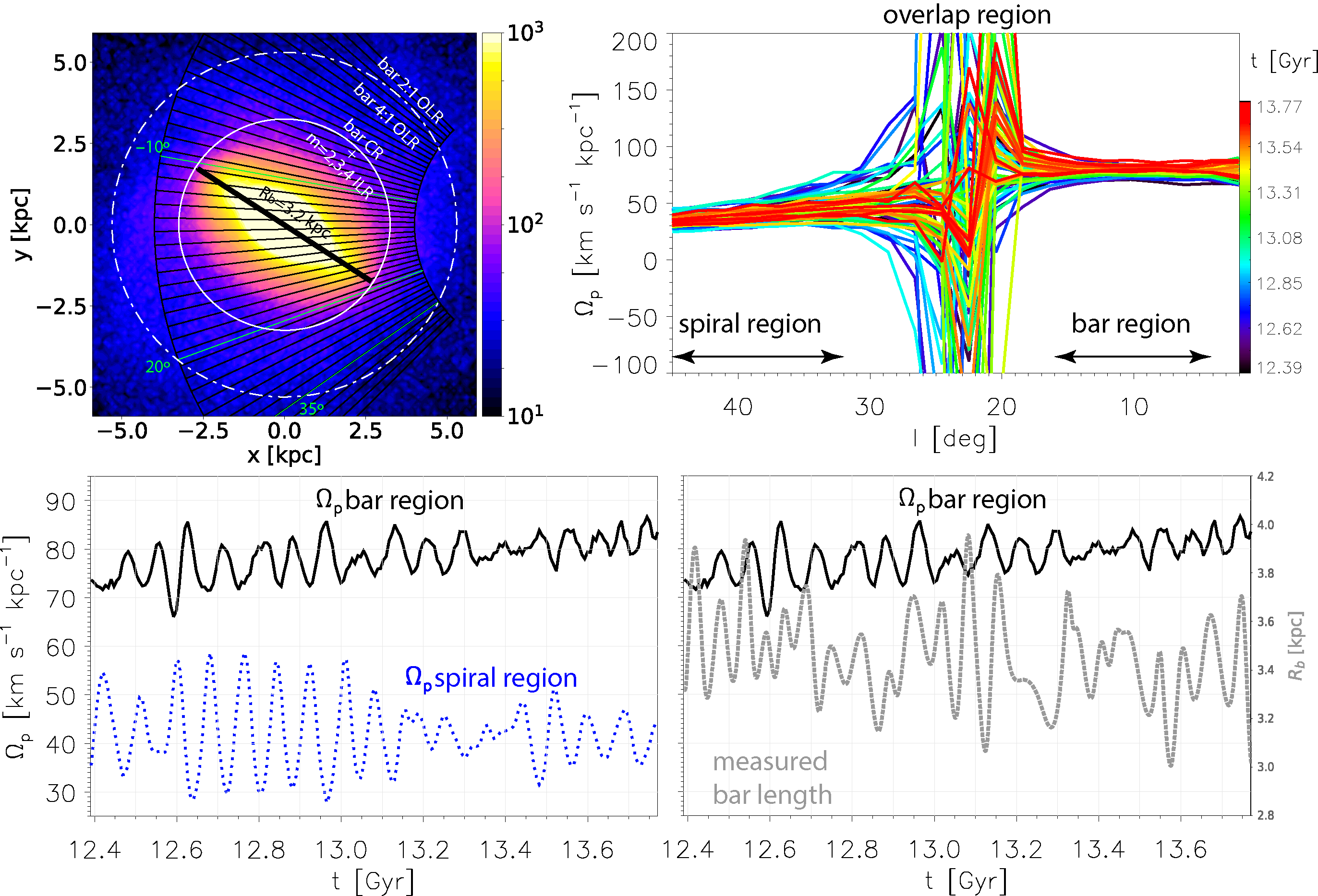}
    \caption{
    Tremaine-Weinberg method applied to Model1 as in the MW, assuming a bar angle of $33^\circ$. {\bf Top left:} Face-on view of the disk stellar density at the final time. Estimates are done in bins of Galactic longitude, $dl=2^\circ$, in the indicated radial range (black arches) and $|b|<5^\circ$. The Sun is at $(x,y)=(8.12, 0)$ and the black line over the bar has a half-length of 3.2~kpc, which corresponds to the time-median $R_{\rm b}$ value extracted from Fig.~\ref{fig:Lcont}.
    The white circles show the bar's time-median CR, 4:1 OLR, and 2:1 OLR at $R_{CR}=3.25$, $R_{4:1OLR}=4.2$, and $R_{2:1OLR}=5.3$~kpc, respectively. The bar's CR radius coincides with the 2:1 ILR of the 2-armed, the 4:1 ILR of a 4-armed, and a 3:1 ILR of a 3-armed spiral, as estimated from Fig.~\ref{fig:power1} (see \S\ref{sec:res}). 
    {\bf Top right:} Estimated $\Omega_\text{p}$ variation with $l$ for the near bar half; color-coded curves correspond to different times, as seen in the color bar. The bar ends just inside $\sim20^\circ$ for our bar angle. The strong divergence at $18\lesssim l\lesssim28$ is caused by the bar-to-spiral transition, which happens between the bar's CR and 4:1 OLR. {\bf Bottom left:} $\Omega_\text{p}(t)$ in the ``spiral region" (dotted-blue curve) and ``bar region" (solid-black curve) obtained by averaging over the longitude ranges indicated in the upper-right panel. The remarkable anti-correlation seen is indicative of bar-spiral mode coupling. {\bf Bottom right:} $\Omega_\text{p}(t)$ in the ``bar region" juxtaposed with the $L_{\rm prof}$ measurement of the corresponding bar half. The $\Omega_\text{p}(t)$ period for both bar and spiral regions is very well defined at $\sim80$~Myr, which lies between the $T_{long}\approx125$~Myr and $T_{short}\approx T_{long}/2$ periods of the measured bar length variation in Fig.~\ref{fig:Lcont}.
    }
    \label{fig:tw1}
\end{figure*}

\begin{figure*}
	\centering
	\includegraphics[width=2.\columnwidth]{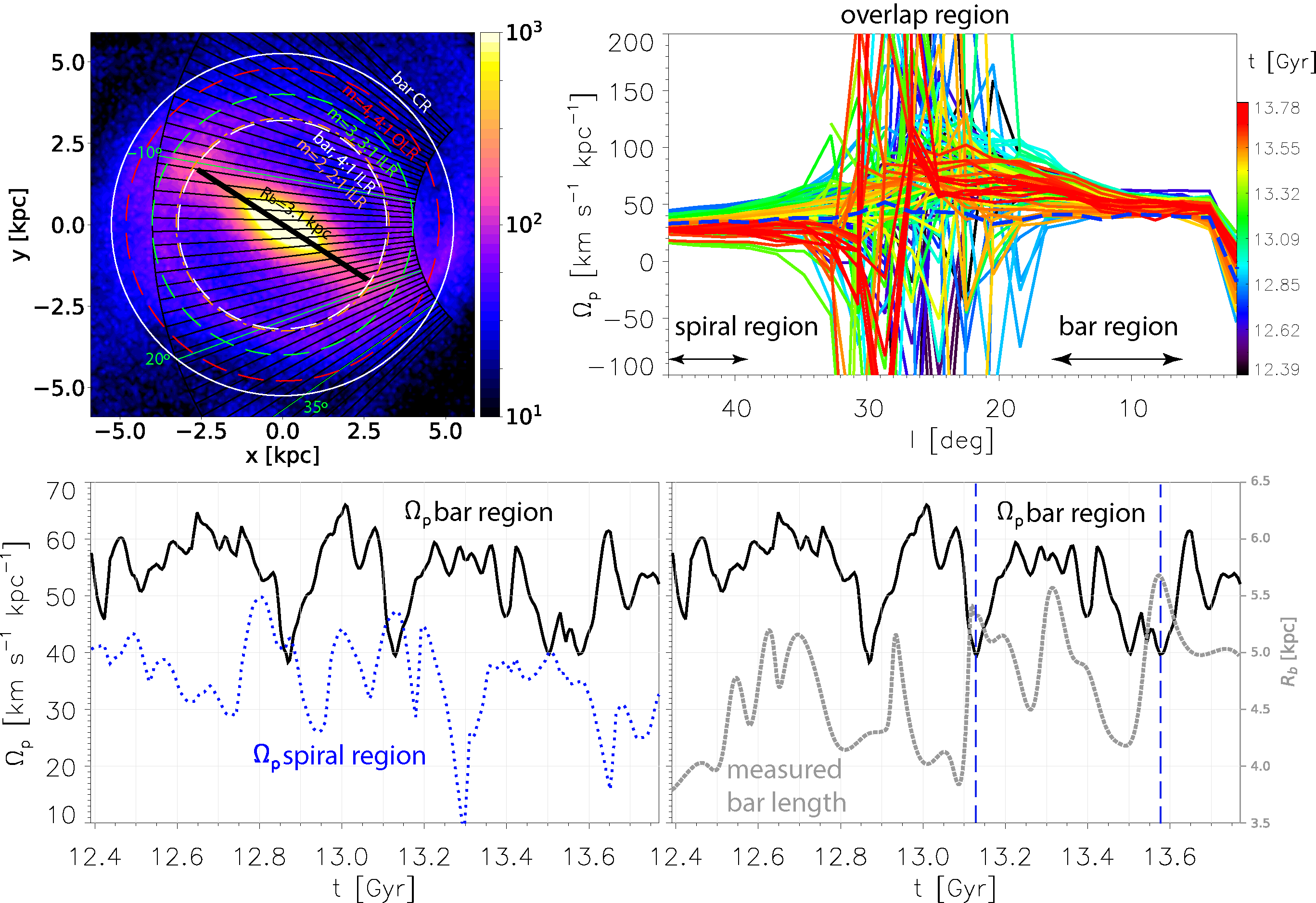}
    \caption{
    As Fig.~\ref{fig:tw1}, but for Model2. The black line has a half-length of 3.1~kpc, indicating the time-median bar length. The white circles show the bar's time-median CR and 4:1 ILR at $R_{CR}=5.25$ and $R_{4:1ILR}=3.2$~kpc, respectively. The orange, green, and red circles show the positions of the 2:1 ILR of a 2-armed, the 3:1 ILR of a 3-armed, and the 4:1 ILR of a 4-armed spiral mode, estimated from power spectrograms (see \S\ref{sec:res}). A ring in the stellar density is seen just outside the bar 4:1 ILR. {\bf Top right:} Estimated $\Omega_\text{p}$ variation with $l$ for the near bar half; color-coded curves correspond to different times, as seen in the color bar. The bar ends just inside $\sim20^\circ$. As for Model1, a strong divergence in $\Omega_\text{p}$ is seen in the transition between bar and spiral regions, with a wider range this time, $17\lesssim l\lesssim33$, due to the stronger spirals.
    The decline of $\Omega_\text{p}$ at $l<6^\circ$ is possibly related to the perpendicular $x_2$ orbits starting to dominate over the bar-supporting $x_1$ orbits. The blue-dashed curve shows a case ($t=13.57$~Gyr), when $\Omega_\text{p}$ is relatively constant out to $l=45^\circ$; this also corresponds to a maximum in the measured bar length (rightmost blue-dashed vertical in bottom-right panel). {\bf Bottom left:} $\Omega_\text{p}(t)$ in the ``spiral region" (dotted-blue curve) and ``bar region" (solid-black curve) obtained by averaging over the longitude ranges in the upper-right panel. A very good anti-correlation is seen, although not as perfect as for Model1, however, the amplitudes are significantly larger. {\bf Bottom right:} $\Omega_\text{p}(t)$ in the ``bar region" is compared with the measured bar length for the positive longitude half (dotted-gray curve, as in Fig.~\ref{fig:Lprof-Martig}). A relatively good anti-correlation can be seen also here (except around $12.6$ and $13.3$~Gyr), with longer bar measurement corresponding to slower $\Omega_\text{p}$. The blue-dashed vertical lines indicate possible configurations for the MW, where the bar appears long ($\sim5.3-5.7$~kpc) and slow ($\Omega_\text{p}\sim40$~\ksk). Note, however, that the average bar pattern speed is $\sim50$~\ksk with variations around the mean of $\sim20\%$, and the true bar length is $\sim3.1$~kpc. 
        }
    \label{fig:tw2}
\end{figure*}

\subsection{Bar amplitude}
\label{sec:strength}

To find out if the bar length fluctuations are accompanied by variations in the bar strength, in Fig.~\ref{fig:FMax-ends} we show the Model1 bar amplitude at five different distances from the disk center along the bar major axis, divided by the maximum, $\eta_R/\eta_{max}$, as indicated, where $\eta\equiv A_2/A_0$ and $\eta_{max}$ is the maximum as a function of radius. As in Fig.~\ref{fig:Methods}, these are estimated from the the $m=2$ Fourier mode, $A_2/A_0$, where $A_0$ is the axisymmetric component.

Very similar time variations appear for all radii, with typical amplitude changes in $\eta_R/\eta_{max}$ of $\sim0.2$, except for the innermost radius considered. The thick solid-gray curve in Fig.~\ref{fig:FMax-ends} represents the mean bar half-length variations (black curve from Fig.~\ref{fig:Lcont2}), which can be seen to agree very well with the amplitude fluctuations, including the short and long periods.

The fluctuations seen in the bar amplitude at fixed radii (Fig.~\ref{fig:FMax-ends}) are more similar to the bar length time variations than the $\eta_{max}$, which can vary with radius (see Fig.~\ref{fig:FMax}). This can generally be seen for Model2 as well in Fig.~\ref{fig:FMax-ends-Marig}. One key difference between the two simulations is the fact that for Model1 all radii peak at nearly the same time, while in Model2 they are delayed with the lowest radius of 3~kpc always peaking last. Such a pattern suggests a spiral arm contribution. Indeed, we established in \S\ref{sec:mod2len} that the Model2 bar true length is about 3~kpc, therefore the region examined in Fig.~\ref{fig:FMax-ends-Marig} lies at or outside the true bar, yet in the regime where our three measurement methods detect a bar. The solid-gray curve in Fig.~\ref{fig:FMax-ends} shows the bar length time variations, seen to follow the overall trend in $\eta_R/\eta_{max}$, in best agreement with the two outermost radii.

\subsection{Bar pattern speed}
\label{sec:omega}

We estimated the {\it instantaneous} bar pattern speed using the modified TW method by \cite{sanders19b}, who applied it to both MW data and N-body simulations. The top-left panel of Fig.~\ref{fig:tw1} shows the configuration we used to measure $\Omega_\text{p}$, over the stellar density of Model1. As done by \cite{sanders19b}, we assumed a bar angle of $33^\circ$, a Galactic latitude range $|b|<5^\circ$, and a solar Galactocentric distance of $R_0=8.12$~kpc. The rotation is clockwise. Estimates are done in bins of Galactic longitude $dl=2^\circ$ in the radial range indicated by the two arches centered on the Sun (at distances of 4.12 and 8.12~kpc), which is sitting at $(x,y)=(8.12, 0)$. The straight black line over the bar has a half-length of 3.2~kpc, which corresponds to $l\sim17^\circ$ for our bar angle. The bar angle is kept the same at each time output. The white circles show the bar time-median CR, 4:1 OLR, and 2:1 OLR at $R_{CR}=3.25$, $R_{4:1OLR}=4.2$, and $R_{2:1OLR}=5.3$~kpc, respectively. The bar CR radius coincides with the 2:1 ILR of a 2-armed, the 4:1 ILR of a 4-armed spiral, and the 3:1 ILR of a 3-armed spiral, estimated from power spectrograms (see \S\ref{sec:power1} and Fig.~\ref{fig:power1} below).

The top-right panel of Fig.~\ref{fig:tw1} shows the estimated $\Omega_\text{p}$ variation with Galactic longitude, $l$, covering the near bar half. The color-coded curves correspond to different times, as indicated in the colorbar. The strong divergence at $18^\circ\lesssim l\lesssim28^\circ$ is caused by the bar-to-spiral transition, happening between the bar CR and 4:1 OLR. To make sense of the pattern speed estimates at different longitudes (and thus, different distances from the Galactic center), we selected a {\it bar-dominated} and a {\it spiral-dominated} regions safely away from the transition region, as indicated by the double-arrows in the top-right panel of Fig.~\ref{fig:tw1}. In the bottom-left panel we plot $\Omega_\text{p}(t)$ for the ``spiral region" (dotted-blue curve) and ``bar region" (solid-black curve) obtained by averaging over the longitude ranges indicated in the upper-right panel. A very good anti-correlation is seen, which is remarkable as these regions are separated by $\sim16^\circ$ in $l$ ($>1.5$~kpc along the bar major axis). This is indicative of a bar-spiral mode coupling \citep{tagger87,quillen11,mf10, petersen19}.
 
In the bottom-right panel of Fig.~\ref{fig:tw1} we juxtaposed $\Omega_\text{p}(t)$ in the ``bar region" with the measured bar length for the positive longitude half (blue-dashed curve in Fig.~\ref{fig:Lprof2}). The $\Omega_\text{p}(t)$ period in both bar and spiral regions is very well defined at $\sim80$~Myr, which lies between the $T_{long}\approx125$~Myr and $T_{short}\approx T_{long}/2$ periods of the measured bar length variation in Fig.~\ref{fig:Lcont}. We relate these frequencies to the coupling between the bar and spiral modes of different multiplicity using power spectrum analyses in \S\ref{sec:power1}.

Fig.~\ref{fig:tw2} is the same as Fig.~\ref{fig:tw1} but for Model2. In the top-left panel, the black line has a half-length of 3.1~kpc, indicating the time-median bar length. The white circles show the bar's time-median CR and 4:1 ILR at $R_{CR}=5.25$ and $R_{ILR}=3.2$~kpc, respectively. The orange, green, and red circles indicate the positions of the 2:1 ILR of a 2-armed, the 3:1 ILR of a 3-armed, and the 4:1 ILR of a 4-armed spiral mode, respectively, estimated from power spectrograms (see Fig.~\ref{fig:power2}). A ring in the stellar density is seen just outside the bar 4:1 ILR. 

As for Model1, a strong divergence in $\Omega_\text{p}$ is seen in the transition between the bar and spiral regions (top-right panel of Fig.~\ref{fig:tw2}), but with a wider range, $17\lesssim l\lesssim33$, because of the stronger spiral arms.
    The decline of $\Omega_\text{p}$ at $l<6^\circ$ is possibly related to the perpendicular $x_2$ orbits starting to dominate over the bar-supporting $x_1$ orbits. The blue-dashed curve shows $t=13.57$~Gyr, when $\Omega_\text{p}$ is relatively constant out to $l=45^\circ$. This also corresponds to a maximum in the measured bar length (rightmost blue-dashed vertical in bottom-right panel). 
    
The bottom-left panel of Fig.~\ref{fig:tw2} shows $\Omega_\text{p}(t)$ in the ``spiral region" (dotted-blue curve) and ``bar region" (solid-black curve) obtained by averaging over the longitude ranges in the upper-right panel. A very good mirror symmetry across the line $\Omega_\text{b}\approx45$ is seen for most peaks, which again we point out as remarkable (as in Model1), since these regions are separated by $\sim24^\circ$ ($\sim2.7$~kpc along the bar major axis). The fractional amplitude of oscillations is significantly larger than for Model1. 

In the bottom-right panel of Fig.~\ref{fig:tw2} we compare $\Omega_\text{p}(t)$ in the ``bar region" to the measured bar length for the positive longitude bar half (dotted-gray curve, as in Fig.~\ref{fig:Lprof-Martig}). A relatively good anti-correlation can be seen also here (except around $12.6$ and $13.3$~Gyr), with longer bar measurement corresponding to slower $\Omega_\text{p}$. The blue-dashed vertical lines indicate possible configurations for the MW, where the bar appears long ($\sim5.3-5.7$~kpc) and slow ($\Omega_\text{p}\sim40$~\ksk). Note, however, that the average bar pattern speed is $\sim50$~\ksk with variations around the mean of $\sim20\%$, and the true bar length is $\sim3.1$~kpc.

The variations of about 20~\ksk seen in the bar region, or $\Omega_\text{p}\approx50\pm10$~\ksk, correspond to a fluctuation around the mean of $\sim20\%$. In addition to four major peaks, one can also see smaller variations on the order of 60~Myr, many of which are also seen in the measured $R_{\rm b}(t)$ (see bottom-right panel of Fig.~\ref{fig:tw2}). We relate these periods of $\Omega_\text{p}(t)$ and $R_{\rm b}(t)$ to the interaction between the bar and spiral modes, using power spectrum analyses in \S\ref{sec:power2}.

More discussion on Fig.~\ref{fig:tw2} is presented in \S\ref{sec:barspirals} and relation to the MW bar is made in \S\ref{sec:MW}.

\section{Evidence for bar-spiral arm interaction}
\label{sec:barspirals}

Since bars typically rotate faster than the spiral structure, there will be times when the two components overlap spatially. As discussed by \cite{comparetta12}, this can be thought of as a constructive interference between two or more waves. \cite{quillen11} noted that in $R-\phi$ density plots the bar seemed to increase in length when connected to the spiral. It is important to consider that spiral structure is never perfectly symmetric in unconstrained simulations, meaning that even 2- or 4-armed spirals will not necessarily connect to the two bar ends at the same time. This is because the density that the bar sees is a combination of the different modes present in the system, which may include $m=1$ and $m=3$ components, as frequently seen in simulations (e.g., \citealt{quillen11, minchev12a}). This can explain why the bar does not grow in length simultaneously on both sides, which is what we found in \S\ref{sec:barsides}. 

It was already evident from Figs.~\ref{fig:tw1} and \ref{fig:tw2} that the bar and spiral are a coupled system for both models, as we showed that their instantaneous pattern speeds fluctuate in time with near-perfect anti-correlation. We explore below a different method of estimating the pattern speeds and search for spiral modes that can explain the fluctuation frequency of our measured bar lengths.

\subsection{Power spectrum analyses}
\label{sec:power}

We constructed power spectrograms using a Fourier transform over a given time window, as described by, e.g., \cite{tagger87}, \cite{masset97}, and \cite{quillen11}.

\begin{table}
\centering
\caption{Fourier modes with corresponding frequencies and pattern speeds for Model1 and Model2, approximated from the spectrograms shown in Figs.~\ref{fig:power1} and \ref{fig:power2}.
}
\label{table1}
\begin{tabular}{lccc}
\hline
&m & $\omega\,\mathrm{[Myr^{-1}]}$ & $\Omega\,\mathrm{[km\,s^{-1}\,kpc^{-1}]}$\\
\hline
Model1 & 1 & 0.03 & 29.3\\
&2 & 0.16 (bar) & 78.2\\
&& 0.05 & 2.4\\
&3 & 0.08 & 26.1\\
&4 & 0.32 & 78.2\\
&& 0.11 & 26.9\\
&& 0.22 & 53.8\\
\hline
Model2 & 1 & 0.08 & 78.2\\
&2 & 0.10 (bar) & 48.9\\
&& 0.04 & 19.6\\
&& 0.07 & 34.2\\
&3 & 0.165 & 53.8\\
&& 0.09 & 29.3\\
&4 & 0.14 & 34.2\\
\hline
\end{tabular}
\end{table}

\subsubsection{Model1}
\label{sec:power1}

\begin{figure}
	\centering
	\includegraphics[width=0.8\columnwidth]{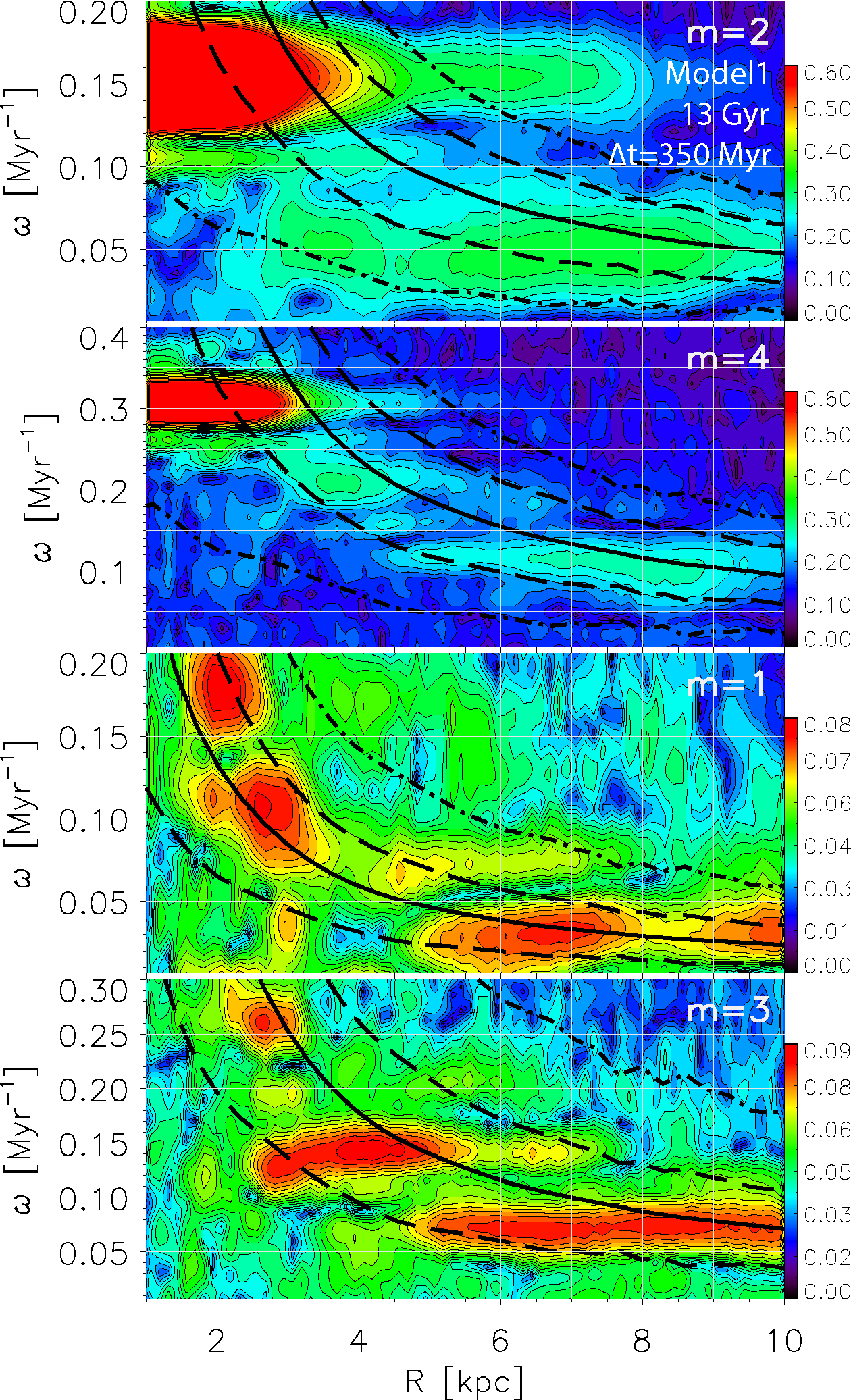}
    \caption{
    Power spectrograms of the $m=1,2,3,$ and 4 Fourier components for Model1 at $t=13$~Gyr using a time window of 350~Myr. The resonance loci are overlaid for the CR (solid-black curve), 4:1 LR (dashed) and 2:1 LR (dot-dashed), computed as $\Omega$, $\Omega\pm\kappa/4$, and $\Omega\pm\kappa/2$, respectively. Contours in the top two panels are saturated at 0.60 (arbitrary scale in color bar) in order to show the spiral structure better. The bar can be seen as the fast red feature in the $m=2$ spectra.
    Most of the modes in the spectrograms can be related to each other as it is expected for a coupled system. For example, the $m=2$ spiral can result from the coupling of the $m=3$ and $m=1$ modes, seen in the range $5<R<10$~kpc, as $\omega_{m2}=\omega_{m3}-\omega_{m1}\approx0.08-0.03=0.05$~Myr$^{-1}$.
    The reconnection period between the bar and the 4-armed spiral at $\omega\approx0.22$ is $T_{rec}=2\pi/\omega_{rec}=2\pi/(4|0.22/4-0.16/2|)\approx63$ Myr and similarly, for the $m=2$ spiral with $\omega\approx0.05$ and the 3-armed mode with $\omega\approx0.14$. The interaction between the bar and these spiral modes (with $m=2,3,$ and 4) thus explains the high-frequency fluctuations ($T_{short}\approx60$~Myr) seen in the wavelength of the low-threshold bar measurements in Fig.~\ref{fig:Lcont}. 
  The longer $R_{\rm b}(t)$ period ($T_{long}\approx125$~Myr) can be related to the $m=1$ mode with $\omega\approx0.03$, which has $T_{rec} \approx2\pi/(2|0.16/2-0.03/1|)\approx63$~Myr, but it meets the same bar half every $\sim126$~Myr. 
    }
    \label{fig:power1}
\end{figure}

In Fig.~\ref{fig:power1} we plot power spectrograms of the $m=1,2,3,$ and 4 Fourier components for Model1 at $t=13$~Gyr using a time window of 350~Myr. The resonance loci are overlaid for the CR (solid-black curve), 4:1 LR (dashed) and 2:1 LR (dot-dashed), computed as $\Omega$, $\Omega\pm\kappa/4$, and $\Omega\pm\kappa/2$, respectively. Contours in the top two panels are saturated at 0.60 (arbitrary scale in color bar) in order to display the spiral structure better.
    
The bar is seen as the red feature in the $m=2$ spectrum (top panel) with $\omega_b\approx0.16$~Myr$^{-1}$. Two slower moving modes (two-armed spirals) are found at $\omega\approx0.05$~Myr$^{-1}$ -- a clump centered near its 2:1 ILR at $\sim3.5$~kpc and the other extending to its CR radius near 10~kpc. Four-armed spirals can be identified in the $m=4$ spectra in the second panel, with the bar's first harmonic seen at $\omega\approx0.32$~Myr$^{-1}$. Clearly defined extended features of relatively constant $\omega$ are found also for the $m=1$ and $m=3$ modes. The rounder clumps of multiplicities 1, 2, and 3 in the inner disk are quite mobile in the vertical direction, when inspecting spectrograms centered on different median times (while the rest are quite stable and long-lived). This is also evident from their vertical extent in this time-averaged plot. An $m=2$ clump seen to shift between the bar and the extended 2-armed spiral as they overlapped was previously reported by \cite{minchev12a}, which is much like what we see in this simulation.

Most of the features found in the spectrograms can be related to each other as it is expected for a coupled system \citep{tagger87,sygnet88, quillen11, minchev12a}. A mode with an azimuthal wave number $m_1$ and frequency $\omega_1$ can couple to another wave with $m_2$ and $\omega_2$ to produce a third one at a beat frequency, $\omega_{beat}$ with the following selection rules:

\begin{equation}\label{eq:m}
m = m_1\pm m_2
\end{equation}
and
\begin{equation}\label{eq:om}
\omega = \omega_1\pm \omega_2
\end{equation}

For example, the low-frequency extended $m=2$ spiral can result from the coupling of the $m=3$ and $m=1$ modes seen in the range $5<R<10$~kpc, as $\omega_{m2}\approx0.08-0.03=0.05$~Myr$^{-1}$ and $m=3-1=2$. Adding these wave numbers gives us an $m=4$ mode with $\omega_{m4}\approx0.11$~Myr$^{-1}$, which is indeed seen in the $m=4$ spectra. Moreover, the bar appears coupled with the 2-armed and the faster 4-armed waves, since $\omega_{m4}-\omega_{m2} \approx 0.22-0.05 = 0.17$~Myr$^{-1}$. The modes and their frequencies used in the following discussion are listed in Table~\ref{table1}.

Which of these modes can explain the measured bar length fluctuations in Fig.~\ref{fig:Lcont}? We can answer this by considering how often the bar encounters different spiral modes.

\begin{figure}
	\centering
	\includegraphics[width=0.8\columnwidth]{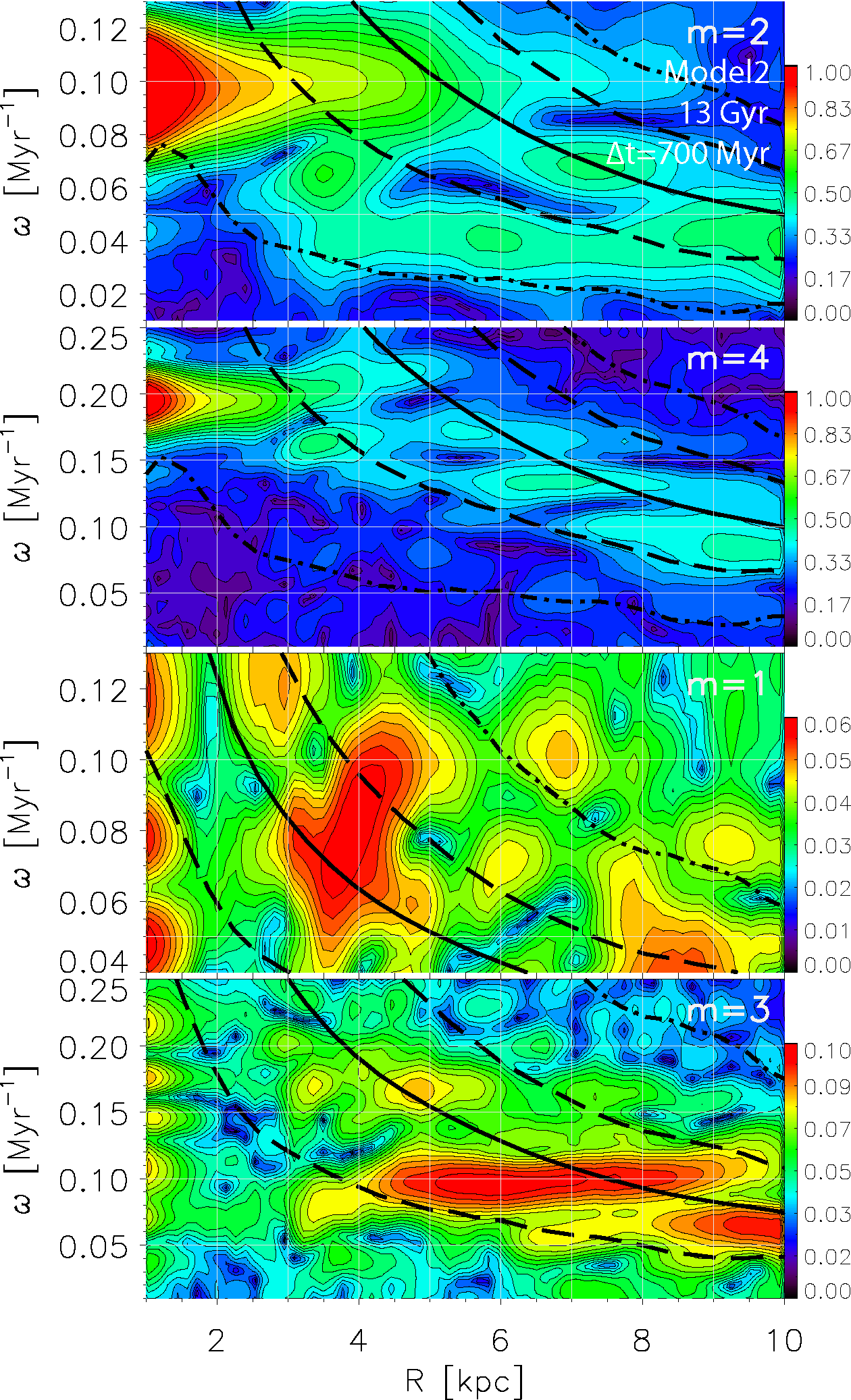}
    \caption{
As Fig.~\ref{fig:power1}, but for Model2. As already seen in the TW-method estimates, this bar is about 50\% slower than Model1's bar. Contours in the top two panels are saturated at 1.0 (arbitrary scale in color bar) in order to show the spiral structure better.
The reconnection period between the bar and the slower 2-armed spiral is $T_{rec}=2\pi/(2|\omega_b/2-\omega_{m2}/2|)=2\pi/|0.1-0.04|\approx105$~Myr. The same reconnection periods with the bar are found for both the $m=4$ spiral with $\omega\approx0.14$ and the $m=3$ mode with $\omega\approx0.09$~Myr$^{-1}$, which can explain the $\sim100$~Myr fluctuations of the $R_{\rm b}(t)$ measurements. As in Model1, $m=2, 3,$ and 4 modes conspire to interact with the bar on the same timescale, here $\sim105$~Myr, suggesting that we have a strongly coupled system. 
The longer period of $200$~Myr in both $R_{\rm b}(t)$ and $\Omega_\text{p}(t)$ can be related to the faster $m=2$ spiral, which has a reconnection period with the bar of $T_{rec}\approx210$~Myr. Even longer reconnection periods result from the $m=1$ mode near $R=4$~kpc and the fast feature in $m=3$ with $\omega\approx0.165$~Myr$^{-1}$, which can be linked to the longer timescale seen in the second half of the time period of Model2 and the wave packet of about $200-400$~Myr found in the TW-method estimated $\Omega_\text{p}(t)$ in Fig.~\ref{fig:tw2}.
    }
    \label{fig:power2}
\end{figure}

\subsubsection{The reconnection frequency, $\omega_{rec}$}

We define here the {\it reconnection} frequency, $\omega_{rec}$, between the bar and a spiral mode, which tell us how often {\it either bar side} is being passed by {\it any spiral arm}. For a spiral with $m=2$, this is $\omega_{rec} = 2|\Omega_\text{b}-\Omega_\text{s}|=|\omega_b-\omega_\text{s}|$. For an $m=4$ mode $\omega_{rec} = 4|\Omega_\text{b}-\Omega_\text{s}|=4|\omega_b/2-\omega_\text{s}/4|$. For two modes of the same multiplicity, therefore, $\omega_{rec}=\omega_{beat}$, where $\omega_{beat}$ is defined as in, e.g., \cite{tagger87}, but this is not the case when the wave numbers are different. For bisymmetric modes, $\omega_{rec}$ gives also the frequency of how often {\it the same bar half} is encountered by {\it any spiral arm}, which is a quantity we are more interested here.

Considering non-bisymmetric modes, for $m=3$ we can write $\omega_{rec} = 6|\Omega_\text{b}-\Omega_\text{s}|$ and for $m=1$ we have $\omega_{rec} = 2|\Omega_\text{b}-\Omega_\text{s}|$. Unlike for even modes, {\it the same bar half} is encountered by {\it any spiral arm} at the frequency $\omega_{rec}/m_b$.

From the above discussion we can write more generally $\omega_{rec} = LCM(m_b,m_s)|\Omega_\text{b}-\Omega_\text{s}|$, where $LCM(m_b,m_s)$ is the Least Common Multiple of the bar and spiral wave numbers. For non-bisymmetric modes we need to divide by the bar wave number to find out how often a given bar half is passed by any spiral arm.

The reconnection period between the bar and the 4-armed spiral at $\omega\approx0.22$~Myr$^{-1}$ is $T_{rec}=2\pi/\omega_{rec}=2\pi/(4|0.22/4-0.16/2|)\approx63$ Myr. Similarly, for the $m=2$ spiral with $\omega\approx0.05$~Myr$^{-1}$ we get $\sim57$~Myr, which is consistent with the $m=4$ mode within our rough estimate of the frequencies. The 3-armed mode just outside the bar with $\omega\approx0.14$~Myr$^{-1}$ has $T_{rec} \approx2\pi/(6|0.16/2-0.14/3|)\approx32$~Myr, but it meets the same bar half every $\sim64$~Myr. The interaction between the bar and these spiral modes (with $m=2,3,$ and 4) thus explains the high-frequency fluctuations ($T_{short}\approx60$~Myr) seen in the wavelength of the low-threshold bar measurements in Fig.~\ref{fig:Lcont}.

We can also explain the longer $R_{\rm b}(t)$ period ($T_{long}\approx125$~Myr) by considering the $m=1$ mode with $\omega\approx0.03$~Myr$^{-1}$, which has $T_{rec} \approx2\pi/(2|0.16/2-0.03/1|)\approx63$~Myr, but it meets the same bar half every $\sim126$~Myr. 

The above modes with $m=1,2,3,$ and 4, including the bar, must be all coupled since they all have $T_{rec}\approx60$~Myr. It appears that the shorter and longer timescales of the bar length fluctuations are related as $T_{short}=T_{long}/2$, resulting from the effect of the slow $m=1$ mode.  

As noted in the discussion of Fig.~\ref{fig:Lcont2}, individual bar halves peak in length at different times, alternating between smaller and larger maxima. This suggests the work of $m=1$ and/or $m=3$ modes, which would naturally connect to each bar side at different times. This departure from bisymmetry can now be explained by the above found $m=1$ mode that we associated with $T_{long}$.

The bar pattern speed resulting from the $m=2$ spectrogram is $\Omega_\text{b}=\omega/m=(0.16/2)\times 977.915\approx78.2$~\ksk, where the last factor fixes the units. We derive a remarkably similar value of $\sim78.5$~\ksk from the TW method applied to the ``bar region" (bottom-left panel of Fig.~\ref{fig:tw1}), after averaging over the 350~Myr time window used for the spectrograms and centered on 13~Gyr.

We can also look in Fig.~\ref{fig:power1} for the spiral whose time-fluctuating pattern speed was measured by the TW method in Fig.~\ref{fig:tw1}. There are two constraints there: (1) we need a mode that has a mean $\Omega_\text{p}\approx40-45$~\ksk and (2) we need the same mode to have a reconnection frequency with the bar of $\sim0.08$~Myr$^{-1}$, in order to explain the $\sim80$~Myr period of $\Omega_\text{p}(t)$ in the ``spiral region" of Fig.~\ref{fig:tw1}. 

These conditions are satisfied for an $m=2$ mode with $\omega_{m2}\sim0.08$~Myr$^{-1}$ or an $m=4$ with $\omega_{m4}\sim0.24$~Myr$^{-1}$, both of which give $\Omega_\text{p}=\omega/m\approx39$~\ksk. Note that these frequencies lie between the bar and the $m=2$ and $m=4$ clumps centered near $R=3.5$~kpc. It may be possible that transient recurring waves shifting back and forth between the bar and slower spirals would result in the strong $\Omega_\text{p}$ fluctuations. Indeed, these two clumps must oscillate in the ranges $0.05<\omega_{m2}<0.12$ and $0.2<\omega_{m4}<0.26$~Myr$^{-1}$, i.e., $25<\Omega_\text{p}<60$~\ksk, which is very much in agreement with the fluctuations in the ``spiral region" of Fig.~\ref{fig:tw1}. For this estimate we considered the bar's lower boundary to be at $\omega\approx0.12$~Myr$^{-1}$, which makes sense since bar minima correspond to spiral maxima in $\Omega_\text{p}$. This can be thought of as the bar speeding up to connect to a spiral arm and slowing down as it disconnects from it (and similarly but opposite for the spiral arm). 

\subsubsection{Model2}
\label{sec:power2}

In Fig.~\ref{fig:power2} we show power spectrograms of Model2 for a time window of 700~Myr, also centered on 13~Gyr. Unlike for Model1 and what is typically seen in other simulations, here we find that the first $m=2$ mode outside the bar is slower ($\omega\approx0.04$~Myr$^{-1}$) than the second one ($\omega\approx0.07$~Myr$^{-1}$), which is probably related to the bar being slow. These 2-armed waves are likely coupled to the bar since the sum of their frequencies is $\sim0.11$~Myr$^{-1}$, which is close to the bar's $\omega\approx0.1$~Myr$^{-1}$. The slow $m=2$ and the faster $m=4$ modes also present evidence for coupling with the bar since $\omega_{m4}-\omega_{m2}\approx0.14-0.04=0.10$~Myr$^{-1}$. A summary of Model2's modes, frequencies, and pattern speeds is given in Table~\ref{table1}.

As discussed about Model1, clumps near the bar end tend to move vertically with time, as the bar and spirals reconnect, e.g., the $m=2$ and $m=4$ features at $R\approx3.5$~kpc. These fluctuations of $\omega$ just outside the bar are likely causing the TW-method $\Omega_\text{p}$ variations with time in Fig.~\ref{fig:tw2}. 

The reconnection period between the bar and the slower 2-armed spiral is $T_{rec}=2\pi/(2|\omega_b/2-\omega_{m2}/2|)=2\pi/|0.1-0.04|\approx105$~Myr. Exactly the same reconnection periods with the bar are found for both the $m=4$ spiral with $\omega\approx0.14$~Myr$^{-1}$ and the strong $m=3$ mode with $\omega\approx0.09$~Myr$^{-1}$. As for Model1, for the 3-armed wave we used $T_{rec}/2$, which gives the frequency of any arm passing the same bar side, which is what is needed here. It is remarkable that, as in Model1, $m=2, 3,$ and 4 modes conspire to interact with the bar on the same timescale of $\sim105$~Myr, assuring us that we have a strongly coupled system. This reconnection period is very close to the $\sim100$~Myr fluctuations of the $R_{\rm b}(t)$ measurements, estimated from the 8 peaks in $L_{\rm prof}$ in the first $\sim800$~Myr (roughly, the time window used for spectrograms) shown in Fig.~\ref{fig:Lprof-all-Martig}; we considered one bar half to account for deviations from bisymmetry.

The faster $m=2$ spiral has a reconnection period with the bar of $T_{rec}\approx210$~Myr. Exactly the same number arises from the $m=1$ mode, if we use $\omega\approx0.08$~Myr$^{-1}$, which is the value that the vertically extended red clump is centered on. We can add a range here by considering the boundaries of this clump, obtaining $140<T_{rec}<420$~Myr. The fast feature in $m=3$ with $\omega\approx0.165$~Myr$^{-1}$ can produce an even longer reconnection period of $315<T_{rec}<630$~Myr. The latter range results from the unknown precise frequency (we considered $0.16<\omega<0.17$~Myr$^{-1}$) as the result is very sensitive to the small denominator in the expression $T_{rec}=2\pi/(3|\omega_b/2-\omega_{m3}/3|)$. The longer periods of these $m=1, 2 ,$ and 3 modes may be responsible for the longer timescale seen in the second half of the Model2 time period and the wave packet of about 200-400~Myr found in the TW-method estimated $\Omega_\text{p}(t)$ (major peaks at $\sim12.45, 12.65, 13.0, 13.3,$ and 3.65~Gyr in the black-solid curve in bottom panels of Fig.~\ref{fig:tw2}).

$T_{rec}\approx105$~Myr found above between the bar and the $m=2, 3,$ and 4 modes is also very close to the short period in $\Omega_\text{p}(t)$ in the TW-method estimates in the bar and spiral regions of Fig.~\ref{fig:tw2}. 

It is clear that a Fourier analyses of the measured bar length and instantaneous pattern speed will extract the individual frequencies contributing to the effect, however, we leave that to future papers.

\subsubsection{Resonances}
\label{sec:res}

For each model, we estimated the positions of the bar's and spiral waves' main resonances in the bar vicinity from the power spectrograms presented in Figs.~\ref{fig:power1} and \ref{fig:power2}. These values were already used in the top-left panels of Figs.~\ref{fig:tw1} and \ref{fig:tw2}, to indicate the radii at which resonances occur in the disks.

For Model1 the bar's time-median CR, 4:1 OLR, and 2:1 OLR are approximately located at $R_{CR}=3.25$, $R_{4:1OLR}=4.2$, and $R_{2:1OLR}=5.3$~kpc, respectively, as estimated from the points at which the resonance loci cross the maximum power of the corresponding feature in the $m=2$ spectrogram of Fig.~\ref{fig:power1}. We can also see that the bar CR radius coincides with the 2:1 ILR of a 2-armed, the 4:1 ILR of a 4-armed, and a 3:1 ILR of a 3-armed spiral mode, all of which are the first order resonances of the corresponding multiplicity wave. These resonances are plotted as white circles on top of the face-on density contours of Model1 in the top-left panel of Fig.~\ref{fig:tw1}.

Similarly, for Model2 we estimate from Fig.~\ref{fig:power2} that for the bar $R_{CR}=5.25$ and $R_{4:1ILR}=3.2$~kpc, respectively. Additionally, we also identify the 2:1 ILR of a 2-armed, the 3:1 ILR of a 3-armed, and the 4:1 ILR of a 4-armed spiral modes, located near $R=3.2, 4.0,$ and 4.8~kpc, respectively. These resonances are plotted as circles on top of the face-on density contours of Model2 in the top-left panel of Fig.~\ref{fig:tw2}.

For both simulations, the regions just outside the bar ends are densely populated with resonances of different patterns and these are exactly the ``overlap regions" where our TW-method pattern speed estimate produces non-sensible results. Indeed, such overlap of resonances can cause non-linear dynamical effects in the region, such as a strong angular momentum exchange \citep{mf10, brunetti11}, and suggests a coupling between the bar and all participating spiral modes \citep{tagger87, sygnet88, masset97}.

\subsection{Phase-space structure near the bar ends}
\label{sec:baruv}

We can also trace the transition from bar to spiral by studying the velocity space structure near the bar ends. Because stars on bar orbits will have different velocities than those affected by a spiral arm, we should see the two types as individual clumps in phase space. 

\begin{figure}[h!]
	\centering
	\includegraphics[width=1.0\columnwidth]{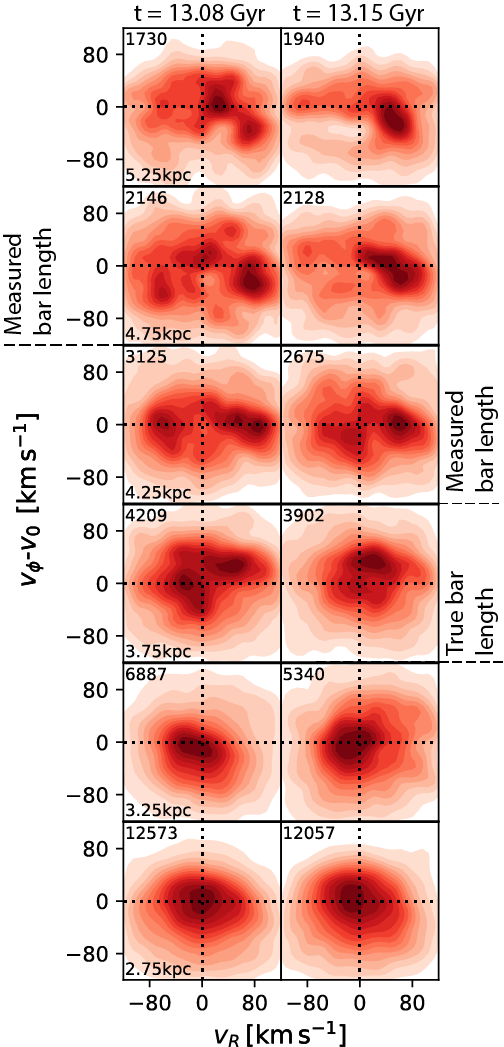}
    \caption{Velocity space along one side of the Model1 bar major axis, for two time outputs (different columns) and six disk neighborhoods along the bar major axis in the range $2.5-5.5$~kpc (median values indicated in the lower-left corners). The local value of the rotation curve is subtracted at each radius. Each neighborhood is 0.5~kpc in radius and 7.5$^{\circ}$ in galactic azimuth. The total number of stars in each spatial bin is shown in the upper left corners. Both snapshots correspond to times when the bar length peaks in Fig.~\ref{fig:Lcont} (blue-dashed curve). The bottom two panels are inside the bar, thus a single clump resulting from the $x_1$ orbits is present. The splitting of the central clump at larger radii ($R=3.75$~kpc and above) indicates that spiral structure extends the bar morphologically to the measured $\sim4.0$ and $\sim4.2$~kpc, although its true length is $\sim3.25$~kpc.
    }
    \label{fig:uv1}
\end{figure}

Fig.~\ref{fig:uv1} shows density plots of radial versus tangential velocity, $V_R-V_\phi$ (known as the $u-v$ plane in the solar vicinity), for two time outputs (different columns) from Model1 and five disk radii along the bar major axis, from 3 to 5.5~kpc. The local rotation curve is subtracted, therefore $V_\phi=0$~\kms corresponds to a tangential velocity equal to the circular velocity of the galaxy at the centre of the bin. Both snapshots correspond to times when the bar length peaks in Fig.~\ref{fig:Lcont}. 

The smallest radius neighbourhoods (bottom panels of Fig.~\ref{fig:uv1}) show one major clump, which can be associated with the $x_1$ bar orbits, as this region is inside the bar, but not too deep to sample the orthogonal $x_2$ orbits (see, e.g., \citealt{contopoulos80a}). As we move to $R=3.75$~kpc another clump appears for both time outputs, likely due to the spiral structure. More clumps appear at larger radii, changing positions with radius, that could be related to different spiral modes.

The single clump seen only in the bottom two panels of Fig.~\ref{fig:uv1} corroborates our conclusion that the true bar length is given by the minimum values measured in Fig.~\ref{fig:Lcont}. The longer bar length measurement at these particular times, however, indicates that spiral structure extends it morphologically to $R\approx4.2$~kpc (see Fig.~\ref{fig:Lcont}). This test can be used to probe the length of the MW bar as more accurate distances and velocities become available in the near future.

\section{Discussion}
\label{sec:discussion}

All three methods used in this work tend to overestimate the bar length, which we interpreted to be due to interaction with the spiral structure. In simulations, for most time outputs visual inspection of the $L_{\rm cont}$ method can reveal artifacts caused by bar-spiral arm connection (see Fig.~\ref{fig:Methods2}); this method has the advantage that each bar half can be examined independently. Such discontinuities will not be seen in the $L_{\rm m=2}$ estimate since the disk is azimuthally averaged. Identifying an artificially large bar measurement with the $L_{\rm prof}$ method appeared to be the hardest.

Interestingly, we found that the $L_{\rm prof}$ method overestimated the Model2 bar for all thresholds (see Fig.~\ref{fig:Lprof-all-Martig}). Even the minimum measurements for the highest threshold overestimate the time-median length by $\sim1$~kpc (or $\sim33\%$). On the other hand, for Model1 $L_{\rm prof}$ produces similar results to the other two methods. 

Unlike $L_{\rm prof}$, the $L_{\rm cont}$ method produces similar results for both models, which is the reason we used the minima in the same threshold to estimate the true bar lengths (see \S\ref{sec:barmethods}; Figs.~\ref{fig:Methods} and \ref{fig:Methods2}).

\subsection{Intrinsic vs apparent bar parameter fluctuations}
\label{sec:intrinsic}

How much of the time fluctuations in bar parameters are simply caused by the constructive interference resulting from the overlapping spiral modes with the bar end, and how much of that is intrinsic? 

As seen from Figs.~\ref{fig:power1} and \ref{fig:power2}, for both of our simulations multiple resonances coexist near the bars ends.  
Such resonance overlap is often seen in N-body simulations and is known to give rise to global non-linear effects (e.g., \citealt{sygnet88, masset97, quillen03, mf10}). If our two models represent coupled systems, as we argued in \S\ref{sec:power}, then it may be expected that intrinsic pulsations in bar parameters may result, due to the non-linear coupling with spiral modes. In such a case, we expect to find fluctuations well inside the bar ``true" length (see \S\ref{sec:mean}). Indeed, in the bar length estimates for Model1 we found that even for thresholds significantly {\it underestimating} the bar length, e.g., the $L_{\rm prof}$ and $L_{\rm m=2}$ methods shown in Figs.~\ref{fig:Lprof} and \ref{fig:Lm}, respectively, fluctuations in $R_{\rm b}(t)$ were still seen. These are always with the longer period, $T_{long}\approx125$~Myr, that we explained with the reconnection frequency between the bar and an $m=1$ mode (see \S\ref{sec:power1}). 

These fluctuations in the Model1 bar length are also accompanied by variations in bar amplitude at small radii, as evident from $\eta_{R=3kpc}$ in Fig.~\ref{fig:FMax-ends}, as well as variations in $\Omega_\text{p}$ well inside the bar (see bottom-left panel of Fig.~\ref{fig:tw1}). All of the above can be seen as evidence of intrinsic variations in these bar parameters, which likely result from the non-linear coupling between the bar and spiral modes. Similar reasoning can be applied to Model2.

On the other hand, there is no doubt that the bar can appear longer when connected to a spiral arm (see, e.g., Fig.~\ref{fig:Methods2}). If the system is coupled, as it is expected, then the intrinsic bar pulsations will coincide with the spiral overlaps, thus the effects will add up. This is likely what we see in Fig.~\ref{fig:Lcont}, where the $L_{\rm cont}$ estimate is shown for different thresholds. High thresholds (e.g., 80\%) that result in shorter length estimates give the true bar fluctuations, since they fall inside the visually-determined bar length. Low thresholds (e.g., 20\%), on the other hand, sample the spiral contribution to the bar length.

We can separate the pattern speed fluctuations in intrinsic and apparent as well. We found time variations in $\Omega_\text{p}(t)$ with an amplitude on the order of 5-10\% (Model1) or 20\% (Model2) by applying the TW method well inside bar region, thus avoiding the overlap with spirals (Fig.~\ref{fig:tw1}).   
In addition to these intrinsic variations, biases in $\Omega_\text{p}$ can arise from a measurement in the bar-spiral overlap region. \cite{wu18} used three different methods to measure the instantaneous $\Omega_\text{p}$ in their simulations, finding a monotonous decrease with radius from $\sim28$ to $\sim12$ \ksk (their Fig.~3), as the bar ($\sim7$~kpc long) transitioned to the spiral. The only explanation for this smooth transition must be that the pattern speed measurement is applied to stars belonging to both the bar and spiral structure. Such worries were already expressed by \cite{rautiainen08}, who warned that some slow bar observations could be caused by this problem. We point out that this is not what we found with our TW-method estimates, since in the bar-to-spiral transition region our estimates diverged strongly.

\subsection{Relation to other theoretical studies}

Most previous works on bars have considered simulations without a gaseous component, where the disk heats quickly rendering spiral structure very weak. Moreover, the high-frequency oscillations with periods as small as 60~Myr we find here can easily be undetected, unless a series of time outputs were analyzed. In contrast to most previous works concerned with bar morphology and evolution, the simulations we used here were hydrodynamical in the cosmological context, even though quiescent in the last stages of evolution that we considered.

Previous evidence for bar parameter fluctuations on a dynamical time scale can be, nevertheless, found in the literature. In their two-dimensional N-body simulations, \cite{rautiainen00} noticed that ``certain relative orientation of bar and spiral modes with different pattern speeds can give a temporary illusion of a considerably longer bar component than the actual one." \cite{quillen11} also reported increase in bar length when connect to spiral arms in their N-body simulations. \cite{martinez11} showed that in their N-body simulation the bar can appear longer by developing a leading end through the interaction with the adjacent spiral arm. Bar length fluctuations can be seen in Figures 11 and 12 by \cite{michel-dansac06}. Bar amplitude variation with time can be seen in Fig.~2 by \cite{minchev12a} for simulation gSb, with a period of $\sim200$~Myr, although the origin of those was not discussed. Similarly, bar amplitude and pattern speed fluctuations with time can be found in Fig.~7 and Fig.~9 by \cite{herpich17}, although not acknowledged in the paper. Pattern speed fluctuations on a dynamical timescale can also be seen in Fig.~7 by \cite{sanders19b}, inferred from the moments of inertia of consecutive snapshots in their N-body simulation. Interestingly, the TW method applied to the same simulation appeared to capture these variations.

\cite{wu18} studied variations in bar length, strength, and pattern speed in a simulation of a double-barred galactic disk, attributing the majority of effects to the inner bar. They concluded that the primary bar was longest when disconnected from the spiral structure. Although they used dissipationless N-body simulations in which the spiral structure must be weak after the first couple of Gyr, this is still the opposite to what we find with our weak spiral structure Model1. Pattern speed and amplitude fluctuations of the secondary bar in double-barred N-body galactic disks was also reported by \cite{debattista07}.

\cite{wu18} noted that in the region between the inner and outer bars there was a disagreement among their three $\Omega_\text{p}$ determination methods, which they attributed to the absence of a well-defined steady pattern. They measured negative pattern speeds reaching values of up to $\Omega_\text{p}\sim-200$~\ksk (their Fig.~3). We can, therefore, expect that the strong scatter in our $\Omega_\text{p}$ estimates in the transitional region between bar and spiral structure (top-right panels of Fig.~\ref{fig:tw1} and \ref{fig:tw2}) results from the same effect.

The simultaneous work of \cite{petersen19} studied the time evolution of bars in N-body simulations, also finding rapid fluctuations in bar length and pattern speed. As for our Model2, they also found strong overestimation of bar length by the ellipse fitting method (our $L_{\rm prof}$).

\subsection{Implications to studies of external galaxies}
\label{sec:external}

Our Model1 can be associated with an SBab galaxy type, owing to its more tightly wound and multi-armed spirals (see Fig.~1 in \citealt{buck18}), while Model2 is more similar to an SBbc type, with its stronger, more open, and dominated by $m=2$ and $m=4$ spiral arms (see top-right panel of Fig.~1 in \citealt{martig14a} or Fig.~1 in \citealt{mcm13}). 

Unlike in simulations, it is not as straightforward to correct for artificially long bar measurements in observations (e.g., CALIFA, MANGA, $\rm S^4G$), where we always see a single snapshot, which may or may not correspond to the bar and spirals being connected. 

One of the most common bar length determinations is ellipse fitting to the isophotes of the galaxy (e.g., \citealt{wozniak91, laine02, sheth03, erwin03}). This is similar to our $L_{\rm prof}$ method, although we apply it to the stellar density in the simulations, which we showed can overestimate the length by up to 30\% in our Model1 (see Fig.~\ref{fig:Lprof2}) and up to 100\% in our Model2 (see Fig.~\ref{fig:Lprof-Martig}).

Because spirals appear connected to the bars in most observations (and thus, most of the time, as also seen in simulations, e.g., \citealt{sellwood88}), it is not easy to establish how much the contribution to the bar length is. The bar would appear the longest when several modes overlap near its end, which is not necessarily obvious in the mass distribution. Fourier image decomposition may be possible to assess this (see, e.g., \citealt{elmegreen92, rix93, henry03}).

The presence of strong spiral structure in the inner disk should be seen as a warning that the bar's length may be overestimated. \cite{diaz-garcia19} found correlation between strong bars and strong spiral arms (and to some degree with the bar length), using 391 nearby galaxies from the $\rm S^4G$ survey. Although we agree with their conclusion that this may result naturally if the disk is unstable to perturbations in general, our results here indicate that we should still expect overestimation caused by overlapping bar and spirals.

Another implication of our results is to the {\it fast} and {\it slow} bar classification, which involves a measure of the pattern speed (as inferred from, e.g., a ring assumed to lie at the CR or a change of sign in the streaming motions of the gas - \citealt{font17}) and the bar length. The usual convention is that a {\it slow bar} is one that has a ratio $\mathscr{R}\equiv R_{CR}/R_{\rm b}>1.4$ and a {\it fast bar} is one with $1.0<\mathscr{R}<1.4$ (e.g., \citealt{athanassoula92}). It should be kept in mind that a source of error in the ratio $\mathscr{R}$ can already result from the bar length and CR radius determination: \cite{michel-dansac06} have shown that $\mathscr{R}$ could increase from 1 to 1.4 just by a change of method.

Because our two bars have almost the same lengths but very different pattern speeds, the bar resonances lie at very different radii for each simulation (see Figs.~\ref{fig:tw1} and \ref{fig:tw2}). Model1 is comparable to the fastest bars found in observations, given by the ratio of the bar's CR radius to its length, $\mathscr{R}\equiv R_{CR}/R_{\rm b}\approx3.1/3.05\approx1.02$, while Model2 is significantly slower with $\mathscr{R}=5.6/3.2\approx1.75$, using the true $R_{\rm b}$ values at the final time. These agree very well with the results of \cite{rautiainen08}, who modeled 38 barred galaxies using optical data from the Ohio State University Bright Spiral Galaxy Survey, finding that $\mathscr{R}$ increases from $1.15\pm0.25$ in types SB0/a-SBab (as Model1) to $1.82\pm0.63$ in SBbc-SBc (as Model2).

The above agreement between observations and our models in the value of $\mathscr{R}$ may appear puzzling since we used the true $R_{\rm b}$ and $R_{CR}$, which is usually not what is measured. Unlike other studies, however, \cite{rautiainen08} determined the bar lengths and pattern speeds by producing a dynamical model for each galaxy and matching the overall disk morphology. This is likely avoiding biases in both $R_{\rm b}$ and $\Omega_\text{p}$. We note also that if both $R_{\rm b}$ and $\Omega_\text{p}$ are biased, but in opposite directions, as we showed is often the case, the ratio $\mathscr{R}$ will remain relatively constant, although we usually find that $\Omega_\text{p}$ does not slow down as much as to account fully for the increase in $R_{\rm b}$. Since variations in $R_{\rm b}$ are much larger for our Model2 (and thus possibly for the SBbc-SBc sample of \citealt{rautiainen08}), this may explain the larger spread in $\mathscr{R}=1.82\pm0.63$.

\subsubsection{Ultra-fast bars}

\cite{aguerri15} reported three CALIFA galaxies (NGC 5205, NGC 5406, and NGC 6497) to host ``ultra-fast" bars (see also \citealt{buta09, guo19}), i.e., $\mathscr{R}<1$, which is in disagreement with theoretical studies, showing that the bar's $x_1$ orbits become unstable beyond CR (e.g., \citealt{contopoulos80b}). The authors considered the possibility that the bar lengths were overestimated but concluded the opposite after a careful visual examination of the images.

In light of our findings, however, where we claim that visual inspection often cannot help, it may be that the relative orientation between the bar and spiral modes in these galaxies is such that it gives the maximum bias to the bar length (i.e., maxima in Fig.~\ref{fig:Lcont} and \ref{fig:Lcont-all-Martig}). Note that this will also correspond to maxima in the strength (see Fig.~\ref{fig:FMax-ends} and \ref{fig:FMax-ends-Marig}). 

It can be seen from Fig.~\ref{fig:Lcont} that the Model1 CR radius (red-dotted line) lies most of the time below the measured $L_{\rm cont}$ length, for the thresholds shown, i.e., the bar appears ultra-fast if biases in $R_{\rm b}$ are considered, but not in $\Omega_\text{p}$. For the $L_{\rm prof}$ and $L_{\rm m=2}$ estimates this happens only for the three thresholds that give the highest bar length estimates in Figs.~\ref{fig:Lprof} and \ref{fig:Lm}, respectively. Considering that \cite{aguerri15} used the TW-method to estimate $\Omega_\text{p}$, a bar appearing longer would generally correspond to a lower pattern speed, thus often cancelling out the effect when the ratio $\mathscr{R}$ is considered. As mentioned above, however, we usually find that $\Omega_\text{p}$ does not decrease as much as to account fully for the increase in $R_{\rm b}$. Therefore, for intrinsically fast bars, there will be configurations that make such galaxies appear to host ultra-fast bars.

\subsubsection{Fraction of overestimated bars in observations of external galaxies}

We can try to estimate how often bars will be significantly overestimated in external galaxies. Using the $L_{\rm prof}$ method from Figs.~\ref{fig:Lprof} and \ref{fig:Lprof-all-Martig}, we can estimate that $\sim50\%$ of the time we would measure a length of $\sim3.7$~kpc for Model1 and $\sim4.7$~kpc for Model2. Using the time-median true bar lengths $\sim3.2$~kpc and $\sim3.0$~kpc, this corresponds to $\sim15\%$ and $\sim55\%$ increase in bar lenght for Model1 and Model2, respectively. 

Assuming the local Universe is ergodic, we expect that the time variations we see in our simulations will correspond to the occurrence in observations. Our results then suggest that in about 50\% of bar measurements of MW-mass external galaxies, the bar lengths of SBab type galaxies (as Model1) are overestimated by $\sim15\%$ and those of SBbc types (as Model2) by $\sim55\%$.

Of course the above estimate will depend on the type of bar measurement and threshold used and can be further refined for individual surveys. In general, we expect that for disks with relatively strong spiral arms, $L_{\rm prof}$ will aways (and for any threshold) measure a bar $\sim30\%$ to $\sim100\%$ longer. The latter comes from the minimum and maximum $L_{\rm prof}$ values in Fig.~\ref{fig:Lprof}.

\subsection{Implications for the MW bar}
\label{sec:MW}

The MW has been considered to be an SBbc type galaxy in the Hubble classification for a long time (e.g., \citealt{kormendy19}), which is similar to our Model2, in that the spiral arms are dominated by two- and four-armed structure and consistent in strength with expectations for the MW (\citealt{siebert12,quillen18a,eilers20},  see \S\ref{sec:sp}).
Recent work using Gaia DR2, however, suggests that Galactic arms may be multiple and tightly wound (e.g., \citealt{quillen18b, donghia19}), which is more like our Model1 (i.e., similar to SBab type).

At the present time, mostly data covering the near MW bar half are available. Even with the best distances available, e.g., VISTA Variables in the Via Lactea (VVV) survey or APOGEE red clump giants, small variations in the density along the bar major axis due to a connected spiral will likely be washed out.
In view of our findings, the direct measurements of a long bar from photometric (VVV, 2MASS, UKIDDS, GLIMPSE) data \citep{wegg15} may in fact be caused by a connected spiral arm and the true bar length may be as small as 3.5-4~kpc (a bit longer than that of Model2). Indeed, \cite{rezaei_kh18} presented an extinction map using red clump and giant stars from the APOGEE DR14, showing the location of the Scutum-Centaurus spiral arm is likely connected to the bar's near side (see their Fig.~4). 

\cite{sanders19b} recently used the TW method adapted to 3D Gaia DR2 data to estimate a MW bar pattern speed of $41\pm3$~\ksk, which is in good agreement with \cite{portail17} and \cite{clarke19}. 
Applying the same method to our simulations, we showed in Figs.~\ref{fig:tw1} and \ref{fig:tw2} that this {\it instantaneous} pattern speed measurement can fluctuate around the mean by $\sim10\%$ and $\sim20\%$ for our fast (Model1) and slow (Model2) bars, respectively. A near perfect anti-correlation between the bars' and spirals' pattern speeds was found for both of our models, which could be explained by the bar speeding up to connect to a spiral arm and slowing down to disconnects from it (and similarly but opposite for the spirals). 

If the MW bar is currently near a maximum due to its being connected to a spiral arm \citep{rezaei_kh18}, then Sanders et al. may have measured an instantaneous $\Omega_\text{p}$ value lower than the mean by up to $\sim20\%$, considering our slow-bar model. This revision, resulting in $\Omega_\text{p} \approx 50$~\ksk, will already be able to explain the difference (within the error) between the direct measurement by \cite{sanders19b} and the more traditional estimate of $53\pm1.5$~\ksk by \cite{mnq07}, based on local kinematics. This difference can result because the TW measurement gives the instantaneous pattern speed, while the velocity field near the Sun (Oort constant C variation with velocity dispersion and Galactic azimuth) most likely reflects the time-averaged $\Omega_\text{b}$.

\cite{bovy19} also estimated the MW bar pattern speed using yet another modification of the TW method, finding similar values to those of \cite{sanders19b} and \cite{portail17}. The authors also tested their method on N-body models from \cite{kawata17} and \cite{hunt13} (Target II and Target IV models), finding good agreement with the ``true" pattern speeds. It is not clear, however, if the time outputs used from those simulations had well-defined spiral structure, which we expect would influence the bar pattern speed measurement. While \cite{kawata17} did not specify the strength of their spirals, \cite{hunt13} reported ``faint spiral structure" for their Target II model. If spiral structure was unimportant in the above simulations, then it is not surprising that \cite{bovy19} found agreement between the true and TW method measured $\Omega_\text{p}$. An inspection of Figs.~2 and 3 by \cite{bovy19}, which show comparison between data and model radial velocity field in the bar region, reveals a more radially concentrated butterfly pattern for the data, which argues that the model bar is longer (and thus slower).

Our application of the TW method, as in \cite{sanders19b}, showed that in the bar-to-spiral transition the different overlapping pattern speeds create a strong divergence in the estimate for both of our models and both bar ends. For a $\sim3.1$~kpc bar oriented at $33^\circ$, this is expected to occur in the range $20^\circ<l<30^\circ$ (top-right panel of Fig.~\ref{fig:tw2}). \cite{sanders19b}, however, only considered the Galactic longitude range $-10^\circ<l<10^\circ$, which lies well inside the bar.

TW-method pattern speed estimates spanning a larger $l-$range may be able to detect the discontinuity in the bar-to-spiral transition, thus informing us on the bar length, as well as the inner spiral structure pattern speed. 

Conversely, the bar-spiral orientation may be such that there is no discontinuity in $\Omega_\text{p}(l)$ for the following reason.
The remarkable anti-correlation between $\Omega_\text{p}$ in the ``bar region" and in the ``spiral region" (see bottom-left panel of Fig.~\ref{fig:tw2}) can be interpreted to first order as the bar's acceleration and spiral deceleration as the two approach each other, and the opposite as they are about to separate. If the bar and dominating inner spiral pattern have relatively similar velocities, as is the case for our slow-bar Model2, there will be times when the bar and spiral will have very similar $\Omega_\text{p}$. These times can be seen in the bottom-left panel of Fig.~\ref{fig:tw2}, when the $\Omega_\text{p}$ in the ``bar region" and ``spiral region" overlap. It can be seen in the right panel that for most of those cases the bar happens to be near a maximum, e.g., the three blue-dashed verticals. The blue-dashed curve in the top-right panel of Fig.~\ref{fig:tw2} shows a configuration, when the variation in $\Omega_\text{p}$ with longitude, $l$, is smooth across the transition region, emulating a long bar. This can then explain the results of \cite{bovy19}, who estimated a relatively constant $\Omega_\text{p}$ out to about $R=5$~kpc (or about $l=31^\circ$ for a bar at their assumed $\phi_b=25^\circ$). 
On the other hand, the drop of $\Omega_{\text p}$ inside $R=4$~kpc in their Fig.~4 can be caused by the ill-defined pattern speed in the transition region between a bar ending inside 4~kpc and the spiral structure just outside it. \cite{chiba19} recently described evidence for a slowing down in the MW bar. We would like to emphasize that this is different from the dynamical timescale effects arising from the bar-spiral interaction described in this work. 

Further work is needed to see what will be the effect of bar parameter fluctuations on the phase-space structure in small disk neighborhoods (i.e., the $u-v$ plane). Due to the large number of particles required, typically such studies are done with test-particle integrations \citep{dehnen00, minchev10, fragkoudi19} or analytically \citep{monari17a,monari19}, in which case the bar is in a steady state. What creates the structure in the $u-v$ plane are the locations of resonances in the disk, which are set by the pattern speed. If the fluctuations are on a dynamical scale slower than a rotation at the solar radius (e.g., as in Model1), it will be, most likely, the time averaged pattern speed that is important.

\section{Conclusions}
\label{sec:conclusions}

In this work we studied how the central bars evolve in the latest evolutionary stages of two simulations in the cosmological context, consistent with key properties of the central MW disk region: one described by \citet{buck18, buck19a, buck19b} (Model1) and the other by \citet{martig09,martig12} (Model2). We applied three different methods of bar length measurements, two well known ($L_{\rm prof}$ and $L_{\rm m=2}$, \citealt{athanassoula02, wegg15, wu18}) and one we developed here ($L_{\rm cont}$), which looked at a drop in the background-subtracted density. The bar strength was measured as a function of time using either the maximum of the Fourier $m=2$ component, $\eta_{max}(t)\equiv max(A_2/A_0)$ (which may vary with radius) or the $\eta_R(t)$ value at a fixed radius, examining different radii. In both cases we found agreement with the time variation of the bar length, with longer bar estimates corresponding to larger amplitudes. The bar pattern speeds were estimated using the modified TW method by \cite{sanders19b}, as recently done for the MW, finding time fluctuations, which for the most part anti-correlated with the bar length and strength.  
Our main findings can be summarized as follows:

$\bullet$ For our Model1, which hosts multi-armed weaker spiral structure, the bar total length $R_{\rm b}$, strength, $\eta_{max}$, and pattern speed, $\Omega_\text{p}$, vary periodically with time by $\sim20\%$, $\sim15\%$, and $\sim10\%$, respectively, due to the interaction with the slower moving spiral modes in the bar vicinity. For our Model2 with stronger spiral arms, $R_{\rm b}$ can be overestimated by up to $100\%$ and $\Omega_\text{p}$ varies around the mean by $\sim20\%$. These fluctuations are on the order of the bar and spiral arm reconnection frequency, with maxima every $\sim60$~Myr for Model1 (fast bar) and $\sim100-200$~Myr for Model2 (slow bar). We believe this is a general phenomenon, which should be found in any dynamically self-consistent barred disk model.

$\bullet$ We found that the bar appears longer and stronger when connected to the spiral structure (see Figs.~\ref{fig:FMax-ends} and \ref{fig:FMax-ends-Marig}). This is caused by two distinct effects that appear to complement each other: (1) intrinsic bar pulsations resulting from the bar-spiral coupling, and (2) the constructive interference from overlapping bar and spiral modes (see \S\ref{sec:intrinsic}).

$\bullet$ Because the two sides of the bar typically do not connect at exactly the same time to a given spiral mode, their individual lengths can oscillate by twice as much as the mean bar length (or $40\%$ for Model1, but less for Model2). 
If the side of the Galactic bar facing us has recently connected to a spiral arm , it could result in an apparent bar length longer by $1- 1.5$~kpc. This is a configuration suggested by the work of \cite{rezaei_kh18}, who found that the Scutum-Centaurus arm is likely adjacent to the bar end in extinction maps using APOGEE DR14. 

$\bullet$ If the near side of the Galactic bar is currently at a maximum, then the far bar half could be significantly shorter. Ongoing and future Galactic surveys, such as APOGEE \citep{majewski17} and 4MOST \citep{dejong12}, will be able to test this.

$\bullet$ 
Using the modified TW method by \cite{sanders19b}, we found that the bar pattern speed fluctuates around the mean by $\sim10\%$ and $\sim20\%$ for our Model1 and Model2, respectively. The latter is enough to account for the difference between $\Omega_\text{b} = 41.3\pm3$~\ksk measured by \cite{sanders19b} in the MW (and very similar values by \citealt{portail17}, \citealt{clarke19}, and \citealt{bovy19}) and the faster estimate of $53\pm1.5$~\ksk by \cite{mnq07}, constrained by the Oort constant C variation with velocity dispersion and Galactic azimuth at the solar radius (see \S\ref{sec:MW}). 
This difference could result because the TW measurement gives the instantaneous pattern speed, while the velocity field near the Sun most likely reflects the time-averaged $\Omega_\text{b}$.
 
$\bullet$ 
Through power spectrum analyses we establish that these bar pulsations, with a period in the range $\sim60-200$~Myr, are caused by its interaction with multiple spiral modes, which are coupled with the bar. These non-axisymmetric mass fluctuations and pattern speed variations introduce a strongly time-dependent potential in the bar vicinity and can be linked to the diffusion of stellar orbits across the bar CR noted in a number of previous works (e.g., \citealt{mf10, brunetti11, minchev11a, dimatteo13}).

$\bullet$
We attempted to separate the effects that cause fluctuations in bar parameters into intrinsic and apparent (\S\ref{sec:intrinsic}). We argued that the former can result from the non-linear coupling between the bar and multiple modes in the disk, while the latter can arise from the overlapping between the bar and spiral densities as a function of time. If the systems are coupled, both intrinsic and apparent effects should be synchronized, which makes it hard to distinguish then from each other. We concluded that the variations in the TW-method derived pattern speed are intrinsic, resulting from the bar-to-spiral reconnection.

$\bullet$ We estimated that in about 50\% of bar measurements in MW-mass external galaxies, the bar lengths of SBab type galaxies (as Model1) are overestimated by $\sim15\%$ and those of SBbc types (as Model2) by $\sim55\%$, depending on the relative orientation between the bar and spiral modes, and the strength of the latter (see \S\ref{sec:external}). Consequently, bars longer than their CR radius reported in the literature, known as ``ultra-fast bars" \cite{aguerri15, buta09}, may simply correspond to the largest biases. Although $\Omega_\text{p}$ typically decreases as the bar grows in size, it may not be sufficient to keep the $\mathscr{R}$ parameter constant. 

$\bullet$ We found a splitting of structure in the $V_R-V_\phi$ plane along the bar major axis of our models when the bar's length was at a maximum. This is another way to confirm that the outer bar morphology results from the overlapping spiral structure. Future Galactic surveys can look for such clumps in velocity space in the vicinity of the near bar end of the MW.

We would like to stress here the necessity of using Galactic models that capture the short-scale dynamical evolution expected in the cosmological context. Considering the perturbative effect of the Sagittarius dwarf galaxy (Sgr) on the MW disk is also very important, as shown by a number of works studying Gaia DR2 data (e.g., \citealt{katz18,antoja18,ramos18,laporte18a,laporte19,bland-hawthorn19}), confirming predictions by \cite{minchev09, quillen09, gomez12a} that the disk was `ringing' while phase-wrapping due to a recent minor merger event.
Indeed, using Gaia DR2 data with distances derived with the StarHorse code \citep{santiago16,queiroz18,anders19}, \cite{carrillo19} showed a reversal in the velocity field near the bar end, which was unlike the expectation from a steady-state bar. The authors found a good match to a simulation by \cite{laporte18a}, which considered the interaction of a Sagittarius-like dSph with the MW, and argued that the bar has been recently strongly perturbed and is currently evolving. 

Future work should consider how bar parameter measurements depend on spiral structure parameters, such as modes of different multiplicities, self-sustained or externally induced. Signatures that can give away the true bar length need to be searched for, for applications to both external galaxies, where the disk global morphology is well seen, and to the MW, where we can study millions of individual stars and detailed chemical abundance information is available. More work is also needed to distinguish between intrinsic bar parameter fluctuations, possibly driven by non-linear mode coupling as we suggested here (see \S\ref{sec:intrinsic}) and apparent variations caused by the constructive interference between bar and spiral modes.

\section*{Acknowledgements}

We acknowledge the Erasmus Programme of the European Union for supporting TH's stay at AIP during the period Feb-Dec 2018, when the large majority of this work was performed. We thank N. Frankel, H. Wozniak, and P. Rautiainen for useful comments and the anonymous referee for a constructive report.
TB gratefully acknowledges the Gauss Centre for Supercomputing e.V. (www.gauss-centre.eu) for funding this project by providing computing time on the GCS Supercomputer SuperMUC at Leibniz Supercomputing Centre (www.lrz.de). This research was carried out on the High Performance Computing resources at New York University Abu Dhabi; Simulations have been performed on the ISAAC cluster of the Max-Planck-Institut f\"ur Astronomie at the Rechenzentrum in Garching and the DRACO cluster at the Rechenzentrum in Garching. We greatly appreciate the contributions of all these computing allocations. TB also acknowledges support by the European Research Council under ERC-CoG grant CRAGSMAN-646955.
BF and GM acknowledge funding from the Agence Nationale de la Recherche (ANR project ANR-18-CE31-0006 and ANR-19-CE31-0017) and from the European Research Council (ERC) under the European Union's Horizon 2020 research and innovation programme (grant agreement No. 834148). 
This work is supported by world premier international research center initiative (WPI), MEXT, Japan.
CW acknowledges funding from the European Union's Horizon 2020 research and innovation program under the Marie Sk\l{}odowska-Curie grant agreement No 798384.
JLS acknowledges the support of the Royal Society.

\bibliographystyle{mnras}
\bibliography{myreferences.bib} 

\appendix

\section{Supplementary figures}

\begin{figure*}
	\centering
	\includegraphics[width=1.8\columnwidth]{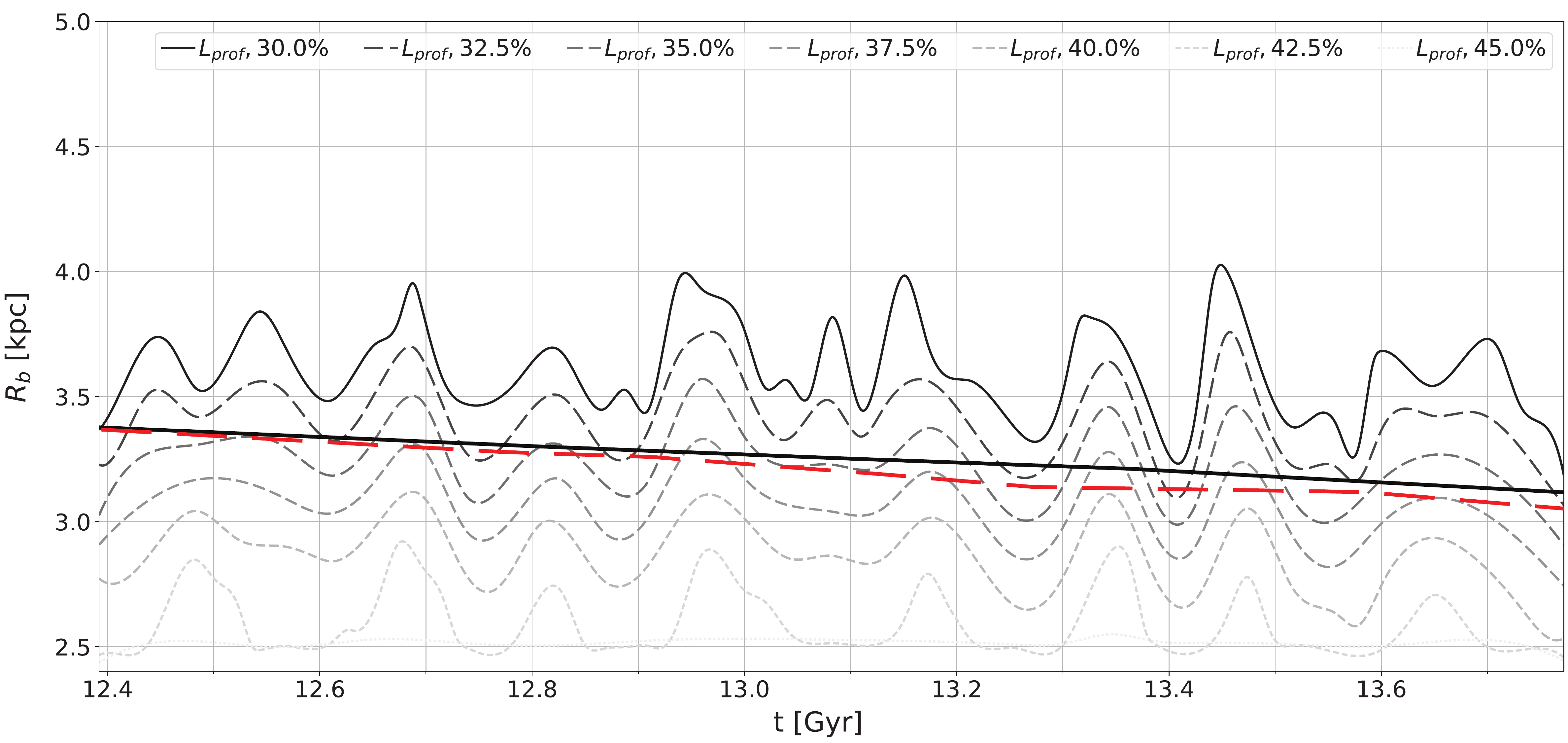}
    \caption[]{As Fig.~\ref{fig:Lcont}, but using the $L_{\rm prof}$ bar length measurement method applied to Model1, as outlined in \S\ref{sec:lprof} and shown in the central column in Fig.~\ref{fig:Methods}. The black-solid and red-dashed lines indicate the variation of the CR radius and the ``true" bar length. The shorter overall bar length resulting from this method, compared to $L_{\rm cont}$, is because we do not consider thresholds smaller than 30\% (note that \citealt{athanassoula02} suggested to use 5\%). Interestingly, the same range of thresholds {\it always} overestimates the bar of Model2 between 30\% and 100\%. 
    }
    \label{fig:Lprof}
\end{figure*}

\begin{figure*}
	\centering
	\includegraphics[width=1.8\columnwidth]{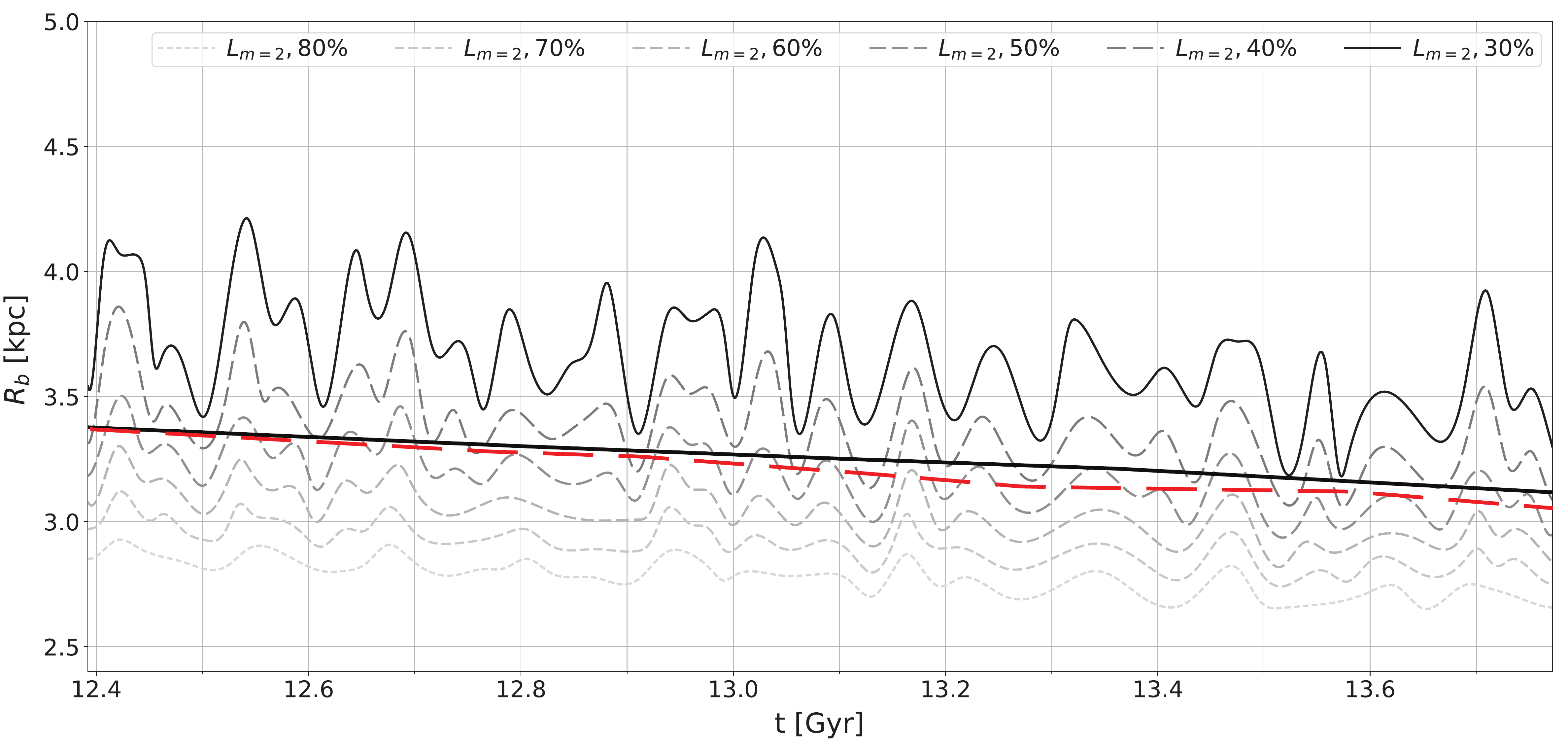}
    \caption[]{As Figs.~\ref{fig:Lcont} and \ref{fig:Lprof}, but measured from the drop in strength of the $m=2$ Fourier component, $L_{\rm m=2}$, applied to Model1. This method is outlined in \S\ref{sec:lm} and shown in the right column in Fig.~\ref{fig:Methods}. The black-solid and red-dashed lines indicate the variation of the CR radius and the ``true" bar length.
    }
    \label{fig:Lm}
\end{figure*}

\begin{figure*}
	\centering
	\includegraphics[width=1.8\columnwidth]{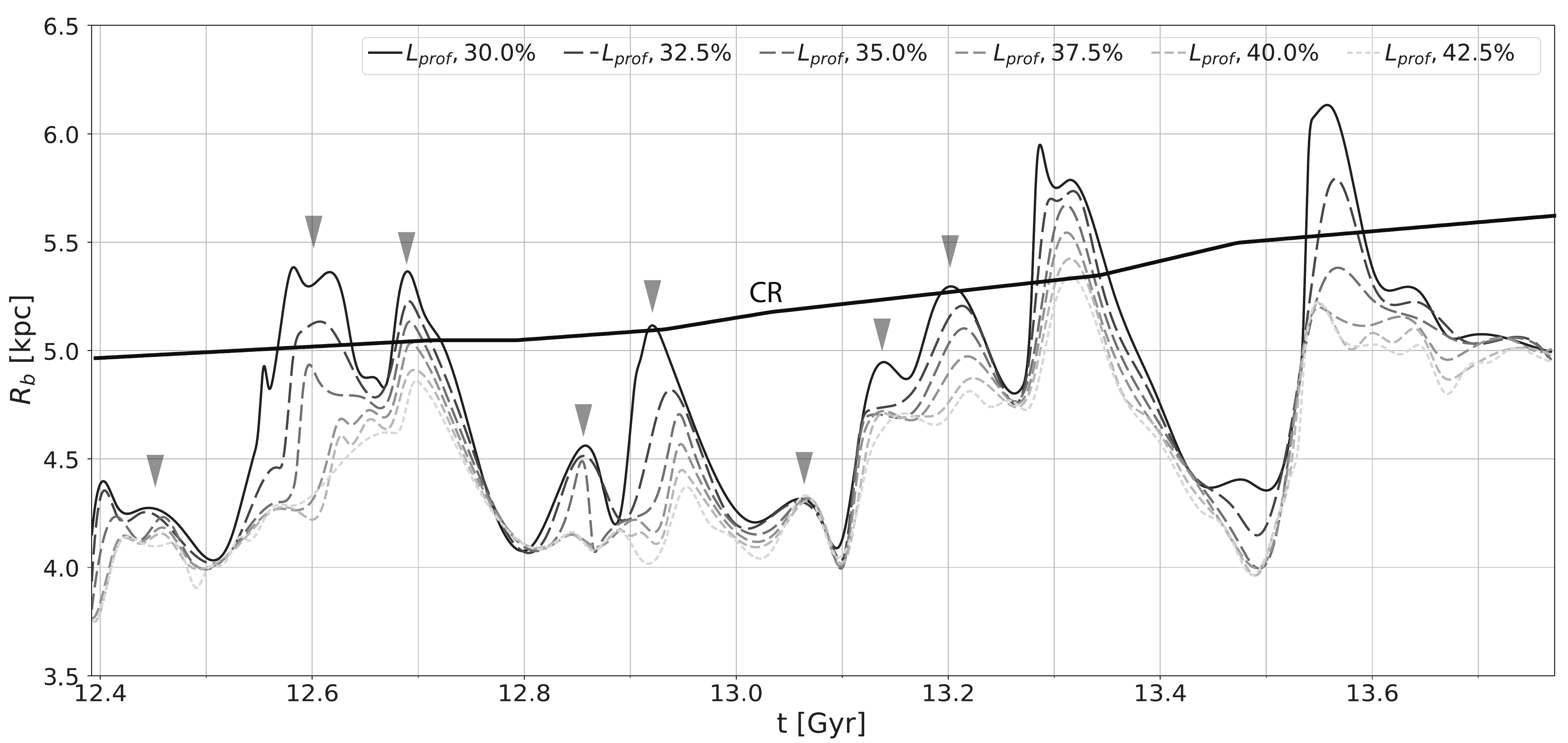}
    \caption[]{
    As Fig.~\ref{fig:Lprof}, but for Model2. The $L_{\rm prof}$ bar length measurement here is much less sensitive to the threshold used, compared to Model1. The true bar length determined by inspecting Fig.~\ref{fig:Methods2} is $\sim30\%$ below the minimum of the highest threshold, thus outside the figure range. At the maximum values measured the overestimating of the bar length is $\sim100\%$. The black-solid line indicates the variation of the CR radius. The gray arrowheads show the main peak positions over a period of 800~Myr used in \S\ref{sec:power2}.
    }
    \label{fig:Lprof-all-Martig}
\end{figure*}

\begin{figure*}
	\includegraphics[width=1.8\columnwidth]{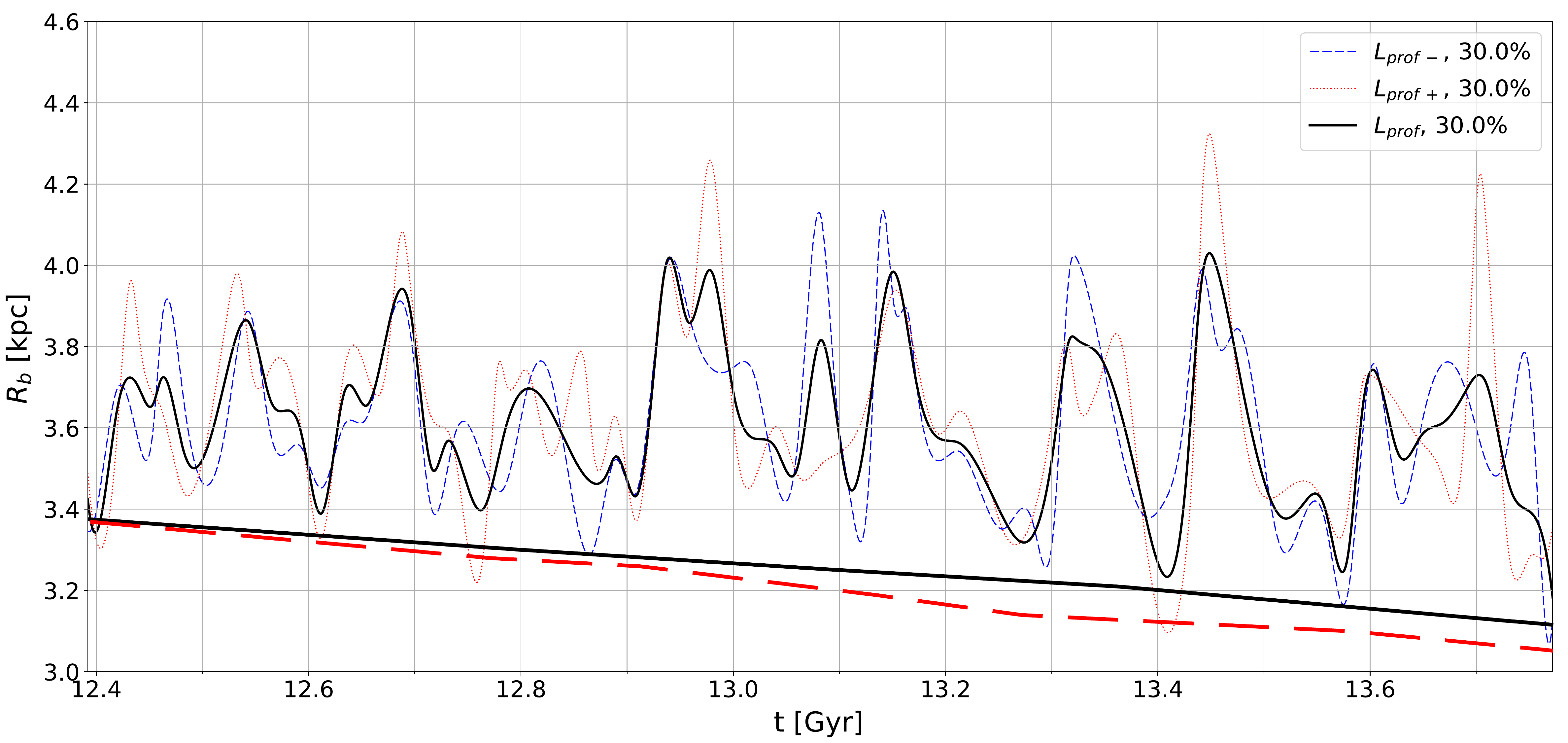}
    \caption{
    Model1 bar length variation with time for individual bar ends, using the $L_{\rm prof}$ method. The black-solid and red-dashed lines indicate the variation of the CR radius and the ``true" bar length. An increase by $\sim40\%$ is seen in the red-dotted curve in the period $13.41<t<13.45$~Gyr.
    }
    \label{fig:Lprof2}
\end{figure*}

\begin{figure*}
	\includegraphics[width=1.8\columnwidth]{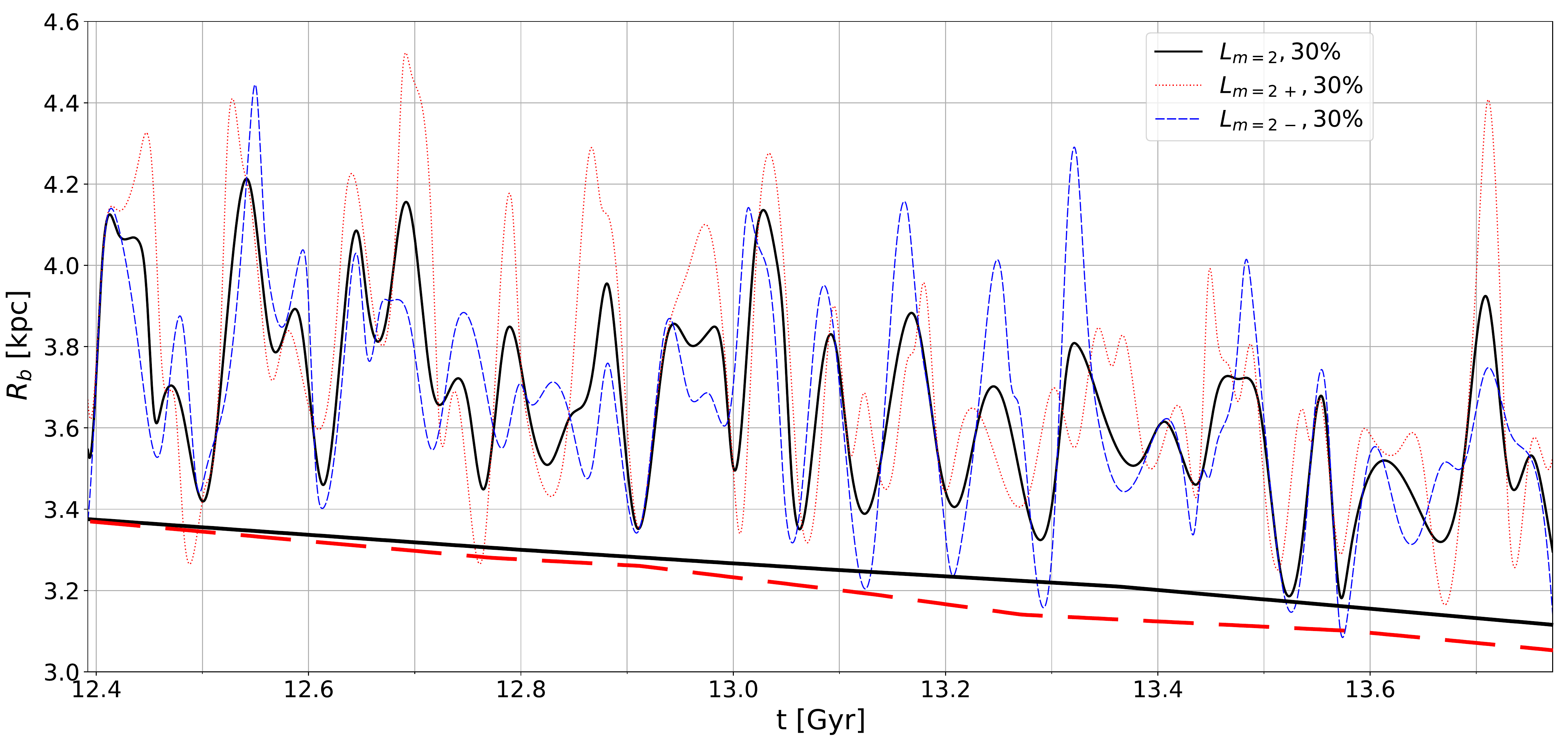}
    \caption{
    Model1 bar length variation with time for individual bar ends, using the $L_{\rm m=2}$ method. The black-solid and red-dashed lines indicate the variation of the CR radius and the ``true" bar length. 
    }
    \label{fig:Lm-30}
\end{figure*}

\begin{figure*}
	\includegraphics[width=1.8\columnwidth]{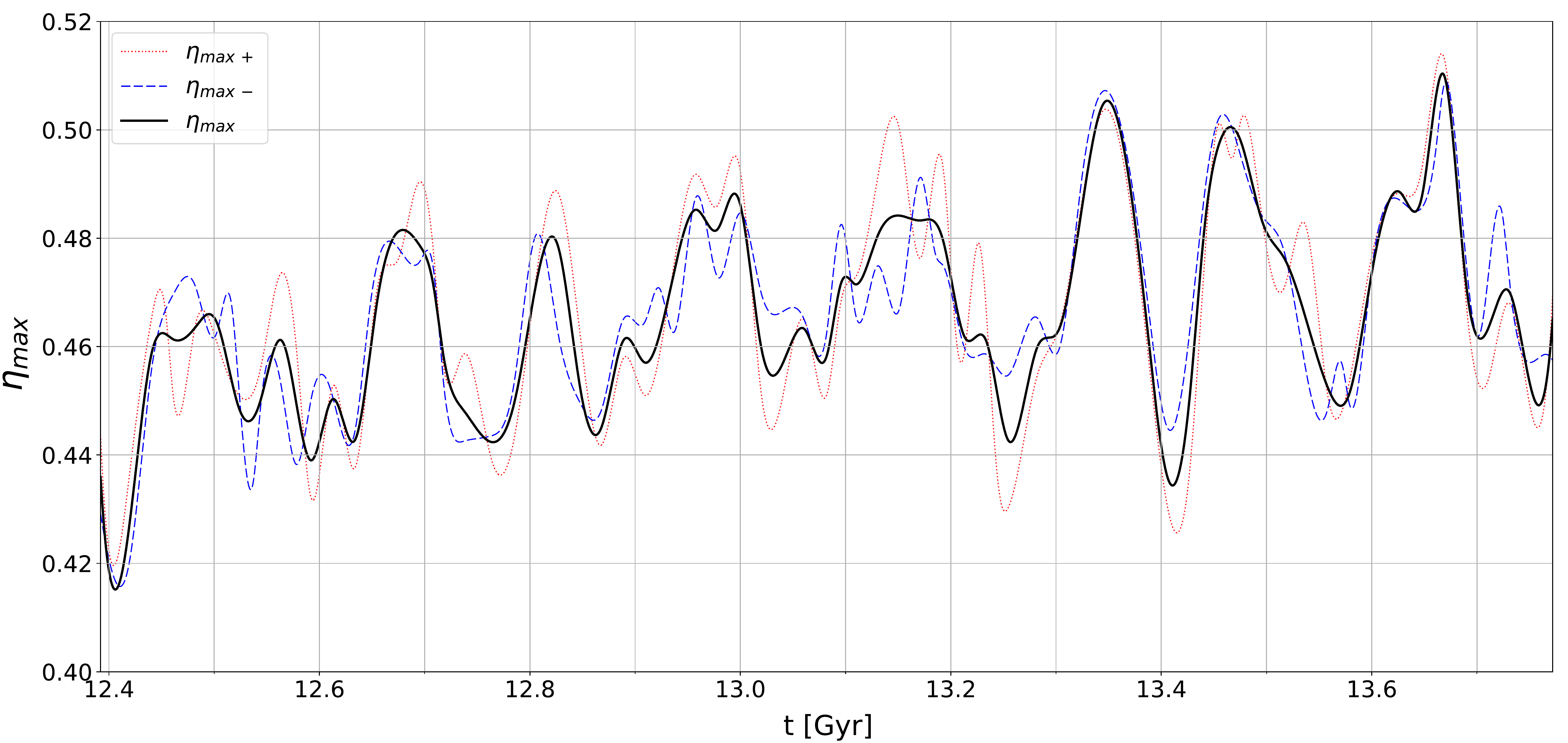}
    \caption{Time evolution of the maximum bar amplitude, $\eta_{max}$, of Model1, estimated from the maximum of the $m=2$ Fourier component, $A_2/A_0$, where $A_0$ is the axisymmetric component (see red triangles in the rightmost column of Fig.~\ref{fig:Methods}). Black-solid curve shows the total component, while red-dotted and blue-dashed curves correspond to the individual bar sides, estimated by reflecting one half of the galaxy across the line $x=0$ prior to taking the Fourier component. Note that the radius at which $\eta_{max}$ occurs can vary; nevertheless, the period is similar to that of the bar length variation with time.
    }
    \label{fig:FMax}
\end{figure*}

\bsp	
\label{lastpage}
\end{document}